\documentstyle[12pt]{article}

\font\ninerm=cmr9

\begin{document}

\title{On the local  systems  Hamiltonian in the
weakly nonlocal Poisson brackets.}

\author{A.Ya. Maltsev$^{(1),(2)}$ \,\, , 
\,\,\, S.P. Novikov\thanks{This work is partially supported by 
the NSF Grant DMS 9704613. It was completed during the stay of 
the author in the Korean Institute for the Advanced Studies 
(KIAS), Seoul.}
$^{(1),(3)}$}

\date{
\ninerm
\centerline{$^{(1)}$ L.D.Landau Institute for Theoretical Physics, 
117940 ul. Kosygina 2, Moscow, }
\centerline{ maltsev@itp.ac.ru \,\, ,
\,\,\, novikov@itp.ac.ru}
\centerline{$^{(2)}$ SISSA-ISAS, Via Beirut 2-4 - 34014 Trieste , 
ITALY}
\centerline{maltsev@sissa.it} 
\centerline{$^{(3)}$ IPST, University of Maryland, 
College Park MD 20742-2431,
USA}
\centerline{novikov@ipst.umd.edu}}

\maketitle

\begin{abstract}

 We study in this work the important class of nonlocal Poisson 
Brackets (PB) which we call weakly nonlocal. They appeared recently 
in some investigations in the Soliton Theory. However  there 
was no theory of such brackets  except very special first order 
case. Even in this case the theory was not developed enough. In
particular, we introduce the Physical forms and find  
Casimirs, Momentum and Canonical forms for the most  important
Hydrodynamic type PB of that kind and their dependence on
the boundary conditions.

\end{abstract}

\centerline{\bf Introduction}

The fundamental idea of the local 
field-theoretical Poisson Brackets (PB)
on the spaces of fields started to  circulate widely in the
community of theoretical and mathematical physicists in the 
second half of the 70s
as a by-product of the KdV theory. It is interesting that
in 1940 L.D.Landau wrote the right formulas
for the local Poisson Brackets of fields in 
hydrodynamics but he called them
"Quantum Commutators"  saying nothing  about Poisson Brackets.
He  realized very soon that he did not need this type 
of quantization of hydrodynamics for the description 
of superfluid $He^4$, and this subject was forgotten
until the late 70s.

In pure mathematics  the idea of symplectic structure
has been considered since the  60s as a most fundamental extension of 
classical Hamiltonian formalism. This point of view is 
certainly true for the geometry of 
finite-dimensional manifolds. However, people  studying PDEs
coming from the problems of
physics  found out soon that the Poisson Structures 
are more fundamental
because they (not the symplectic structures) are local
in most important cases. Already the first (Gardner-Zakharov-Faddeev's)
Poisson structure for KdV (1971) is a  nonstandard local PB; 
however, it was originally described
in the less convenient terminology as a "nonlocal symplectic structure"
\cite{ZF}, or even in the standard trigonometric canonical coordinates
for  periodic boundary conditions \cite{G}.
 Hamiltonian formalism became very important in the development of
 KdV theory: for example, it played
an extremely important role   in  Novikov's approach to the
 solution of the periodic problem   through  stationary higher KdVs
 starting
from the work \cite{N} in 1974--see also the survey article \cite{DMN}.

 The second (Lenard-Magri) local Poisson Structure
for KdV and the idea of $\lambda$-pencils of  compatible PB
structures were
 discovered in 1977 (see in the book \cite{fadtah}).
 After that a huge number of nontrivial local PB
 appeared in the  problems of mathematical physics especially
 for the different classes of  Integrable Systems. 
Another fundamental class of 
 local "hydrodynamic type" PB was discovered in 1983 (see \cite{dn1}
 and Section 3)
 describing the Hamiltonian formalism of the first order quasilinear
 (i.e. "hydrodynamic type") systems in terms of Riemannian Geometry.
 Some specific nonlocal extension of this class started in the works
 \cite{mohfer1,fer1} plays an important role in the theory of
  hydrodynamic type PD systems.
 We call such PB a {\bf Weakly Nonlocal 
Hydrodynamic Type PB}--see Section 3.
 The main part of this work is dedicated to the study of such brackets.
 In particular, in the present work we found all sets of Casimirs and
 canonical  forms for them; we clarified  their dependence
 on the boundary conditions, which is very important in the 
nonlocal case (see  sections 4 and 5).

 As it was observed already in the late 70s, the compatible pair
 of two  Poisson Structures leads to the ''Recursion Operator''
 (see  Section 2 below). This recursion operator produces
  an infinite series of nonlocal
 Poisson Structures: therefore, the KdV equation, for example,
   is the Hamiltonian system
 relative to the  infinite number of   nonlocal Poisson Structures
 with different Kruskal integrals as the Hamiltonians.
The structure of these
 nonlocalities was not clarified until the 90s. In the case of KdV the
 right formula for them
 was obtained in the
 work \cite{enrubor} in 1993. The case of NLS is more complicated:
 there is only one
 local PB for it, all others are nonlocal. In the present work we
 clarified the structure of their nonlocality for NLS (see Section 2).
 As a by-product of this result we introduce a general notion of the
 {\bf Weakly Nonlocal Poisson Bracket.}
 This notion is very natural for the (1+1)-systems. Such brackets
 produce local PD systems for the broad classes of local Hamiltonians.
 They appear in many integrable systems but can describe also a lot of
 non-integrable local perturbations. As it was found recently by Maltsev,
 this class of Poisson Brackets is closed under the operation of the
 "Whitham Averaging"  (it is a nonlinear analog of the
 WKB approximation; people frequently call it a "method of the slow
  modulations of parameters"). This method is
 based on the proper family of quasiperiodic solutions (invariant tori):

  The slow modulation of parameters
always leads  to the hydrodynamic type system Hamiltonian in the
 weakly nonlocal hydrodynamic type PB if you started from the local 
evolution system
 generated by the local Hamiltonian in any weakly nonlocal PB (see
 \cite{malnloc1,malnloc2}).
For the local case this theory was developed by Dubrovin and Novikov
 in 1983 \cite{dn1} (see also the survey article \cite{dn2}). 
However, in 1992 a gap was found in the general 
proof of the Jacoby Identity for the
 hydrodynamic type PB constructed in these works (see \cite{novmal}).
  This gap was
  fulfilled by Maltsev in 1998: full proof can be found in the work
  \cite{engam}.

Our results presented here can be divided into two parts:

Part I (see the Section 2)  
explains how Weakly Nonlocal Poisson Brackets
and Symplectic Structures  
appear from the theory of the famous completely
integrable soliton systems like KdV and NLS. It turns out that all
Higher Poisson Brackets (known since the late 70s)
are weakly nonlocal  for  $n\geq 0$.
For the case of KdV this result follows 
from the work \cite{enrubor}, for NLS
it is completely new. We prove also that all
Higher Symplectic Structures are weakly nonlocal for $n\leq 0$.
 This result is new
for the both cases.

Part II (see the Sections 3--5) is dedicated to the Weakly Nonlocal
 Poisson Brackets
of Hydrodynamic Type associated with Riemannian Geometry. 
 The most important
results known before were obtained by Ferapontov, Mokhov and Pavlov
(see ~\cite{mohfer1}, ~\cite{fer4}, ~\cite{pavlov2}). We formulate
their results below.
 In this work we found the Canonical forms, Casimirs and
the Hamiltonians for the Structure Flows 
(proving therefore that they are the
Hamiltonian Systems). It turns out that these important
 quantities depend on the boundary conditions. We prove also that
 the Symplectic Structure is weakly 
nonlocal for all such Poisson Brackets.

\vspace{0.3cm}

\section{Local and Weakly Nonlocal Poisson and Symplectic Structures.}
\setcounter{equation}{0}

\vspace{0.3cm}

 We are going to consider only one-dimensional (i.e. 1+1) systems, 
so we define
the Poisson Structures either on the spaces of loops
$L({\cal M}^N)$ containing the mappings
$f:S^1\rightarrow {\cal M}^N$ of the circle (periodic boundary conditions)

or on the spaces $L({\cal M}^N,y)$ of  mappings of the line
$R\rightarrow {\cal M}^N$
 constant at infinity,
into some manifold ${\cal M}^N$ with local coordinates
$\varphi^1,\ldots \varphi^n$ where $\varphi(y)=0$. Therefore we  
think about these mappings as vector-functions (fields) 
$\varphi(x)=(\varphi^1(x),\ldots \varphi^n(x))$.
In the nonlocal case we consider only 
  the boundary conditions constant at infinity, i.e.  our mapping
 should be such
that for $x\rightarrow\pm\infty$ we have
$f(x)\rightarrow y\in {\cal M}^N$ and $\varphi(x)\rightarrow 0$.
Here $y$ is some point of the manifold ${\cal M}^N$.
We require also that all derivatives of the map $f$ tend to zero at
infinity. 

 The local field-theoretical PB can be completely defined by the finite
 order differential
 Hamiltonian operator of the form:
$$J^{ij} = \sum_{L\geq k\geq 0}
B^{ij}_{k}(x,\varphi(x),\varphi_{x}(x),\dots) {\partial_x}^{k}$$
                                                                        
  We call this PB translation invariant if the Hamiltonian operator
does not depend on $x$. The  Hamiltonian system generated by the
Hamiltonian functional $I\{\varphi\}$ has the form
$$\varphi^i_t=J^{ij}{\frac{\delta I\{\varphi\}}{\delta \varphi^j(x)}}$$

 The Poisson bracket of two local functionals is
 defined by the formula
$$\{I_{1},I_{2}\}=\int {\delta I_{1} \over \delta \varphi^{i}(x)} 
J^{ij} {\delta I_{2} \over \delta \varphi^{j}(x)} dx$$                  

 In the field theory people prefer 
to write the Poisson Brackets for
fields in the local form convenient for calculations
$$\{\varphi^i(x),\varphi^j(y)\}=J^{ij}\delta (x-y)$$

These formulas define a skew-symmetric bilinear operation satisfying
 to the so-called Leibnitz Identity $\{fg,h\}=f\{g,h\}+g\{f,h\}$.
 It should satisfy also to the Jacoby Identity   for 3 functionals:
 $$\{\{f,g\},h\}+\{\{h,f\},g\}+\{\{g,h\},f\}=0$$

This requirement is very strong. It leads to specific,
very serious restrictions on
the class of admissible Hamiltonian operators.

We define {\bf the weakly non-local Poisson brackets (PB)} 
through the corresponding  class of Hamiltonian operators:

\begin{equation}
\label{jform}
J^{ij} = \sum_{k\geq 0}
B^{ij}_{k}(\varphi,\varphi_{x},\dots) \partial_{x}^{k} 
+ \sum_{k,l \geq 0} e_{kl} S^{i}_{(k)}(\varphi,\varphi_{x},\dots)
{\partial}^{-1} S^{j}_{(l)}(\varphi,\varphi_{x},\dots)
\end{equation}
where $\partial_x\equiv d/dx$ and ${\partial}^{-1}$ is defined
here as a skew-symmetric operator on the line

\begin{equation}
\label{d1}
{\partial}^{-1} = {1 \over 2} \int_{-\infty}^{x} dx -
{1 \over 2} \int_{x}^{+\infty} dx 
\end{equation}
on the space of rapidly decreasing at $\pm\infty$ vector-functions.
The constants $e_{kl} = e_{lk}$ give a quadratic form in the
linear space generated by the (linearly independent) flows
$S_{(k)}(\varphi,\varphi_{x},\dots)$. We say that the bracket
is written in the {\bf Reduced Form} if $e_{kl} = e_{k}\delta_{kl}$
where $e_{k} = \pm 1$.

 Let us point out that we define only translation invariant weakly
nonlocal Hamiltonian operators. It is no problem to extend 
this definition to the
coefficients dependent on $x$, but such brackets will not be considered
in this work.

 In the same way we define a class of weakly-nonlocal 
{\bf Symplectic Structures}. We call the symplectic structure
weakly-nonlocal if the operator $J^{-1}$ (if it exists) has a 
weakly-nonlocal form, i.e.

\begin{equation}
\label{wnsymp}
\left( J^{-1} \right)_{ij} = \sum_{k=0}^{N}
C_{(k)ij}(\varphi,\varphi_{x},\dots) \partial^{k} +
\sum_{k,l \geq 0} d_{kl} Q_{(k)i}(\varphi,\varphi_{x},\dots)
{\partial}^{-1} Q_{(l)j}(\varphi,\varphi_{x},\dots)
\end{equation}

 As far as we know, the first example of Poisson bracket written
in the literature precisely in this form was
the Sokolov bracket (~\cite{sokolov}) 

\begin{equation}
\label{sokolov}
\{\varphi(x), \varphi(y)\} = 
\varphi_{x} \nu (x-y) \varphi_{y} 
\end{equation}
designed to prove that  the so-called Krichever-Novikov equation
is Hamiltonian:

$$\varphi_{t} = \varphi_{xxx} - {3 \over 2}
{\varphi_{xx}^{2} \over \varphi_{x}} +
{h(\varphi) \over \varphi_{x}} =
\varphi_{x} {\partial}^{-1} \varphi_{x}
{\delta H \over \delta \varphi_{x}} $$
where     
$h(\varphi) = c_{3}\varphi^{3} + c_{2}\varphi^{2} +
c_{1}\varphi + c_{0}$ and

$$H = \int \left( {1 \over 2} 
{\varphi_{xx}^{2} \over \varphi_{x}^{2}}
+ {1 \over 3} {h(\varphi) \over \varphi_{x}^{2}} 
\right) dx $$

 For the Sokolov bracket the corresponding symplectic
structure is local:

$$J^{-1} = {1 \over \varphi_{x}} \partial {1 \over \varphi_{x}}$$

 This equation appeared originally in  work ~\cite{KN80} 
describing the "rank 2" solutions of the KP system. 
In  pure algebra it describes  the deformations of the commuting
genus 1 pairs  OD operators of the rank 2 whose classification
was obtained in this work. As it was found later, 
this equation is a unique
third order in $x$  completely integrable evolution equation which
cannot be reduced to KdV by 
Miura type transformations.

 Let us mention that the  
local symplectic structures was considered
by I.Dorfman and O.I.Mokhov (see Review ~\cite{mohrev}).
  
 For the Weakly Nonlocal Poisson Brackets of basic fields we use
the same definition as above. They have  the following form:

$$\{\varphi^{i}(x), \varphi^{j}(y)\} =J^{ij}\delta(x-y)= 
\sum_{k\geq 0}
B^{ij}_{k}(\varphi,\varphi_{x},\dots) \delta^{(k)}(x-y) + $$
\begin{equation}
\label{genbr}
+ \sum_{k,l} e_{kl} S^{i}_{(k)}(\varphi,\varphi_{x},\dots)
\nu (x-y) S^{j}_{(l)}(\varphi,\varphi_{y},\dots)
\end{equation}
where $e_{kl} =e_{lk}$ is a constant symmetric matrix, 
$\delta^{(k)}(x-y) \equiv d^{k}/dx^{k} \delta(x-y)$, 
$\nu (x-y) = 1/2 \,\,sgn(x-y)$ and both sums contain the 
finite  number of terms. All functions involved depend on 
the finite numbers of derivatives of $\varphi$ with respect 
to $x$. The bracket is written in the Reduced form if 
$e_{kl} = e_{k}\delta_{kl}$ and $e_{k} = \pm 1$.

 We include in the definition the following requirements:
all flows with right-hand parts equal to 
$S^{i}_{(k)}, k=1,2,\ldots $
form a linearly independent set; the flows
\begin{equation}
\label{canfl}
{\varphi}^{i}_t = S^{i}_{(k)} (\varphi,\varphi_{x},\dots)
\end{equation}
commute with each other. We call them the
{\bf Structure Flows} for a given weakly-nonlocal
Hamiltonian structure.
The structure flows should preserve the Poisson
structure (\ref{genbr}). Both requirements were proved in the
works \cite{malnloc1, malnloc2} as a corollary from the
definition of the weakly nonlocal PB given above, but here we
simply include them in the definition.

 The brackets (\ref{genbr}) were already used in fact in the recent
 works ~\cite{malnloc1} and
~\cite{malnloc2} where the nonlocal Hamiltonian 
version of the Whitham averaging method
was considered for the local  systems (PDEs) with 
the full necessary set of local commuting integrals. The Poisson
  Brackets used in these works are in fact weakly nonlocal.

\vspace{0.5cm}

 We call the weakly nonlocal Poisson Bracket {\bf fundamental}
if it contains  only one flow  of the
form $S^i_0=\varphi^i_x$ in the non-local part (i.e. its nonlocal part
exactly coincides with the Sokolov PB written above).

In this case 
{\bf every local translation invariant Hamiltonian}

$$H=\int h(\varphi(x),\varphi_x(x),\ldots, \varphi_{x...x}(x))dx$$
{\bf generates the local (PDE) system} because:
$$\varphi^{i}_{x}  {\delta H \over \delta \varphi^{i}(x)} = \partial_x
Q(\varphi, \varphi_{x},\ldots )$$
for some function $Q$.

  For the more general weakly nonlocal
Poisson brackets defined above this property is valid only for 
the special Hamiltonians:  it was pointed out by E.V.Ferapontov  
(\cite{fer3}), that for the nonlocal brackets of Hydrodynamic Type 
 
{\bf local Hamiltonian generates local system if and only if
it is a conservative quantity for the flows $\varphi^{i}=S^{i}_{(k)}$
in the nonlocal part
of the bracket.}
 
This statement is valid for
any weakly nonlocal bracket, and the proof is the same.

 Let us mention also that it is easy to check by direct calculation 
that for any closed 2-form (\ref{wnsymp}) the corresponding forms
$Q_{(k)i}(\varphi,\varphi_{x},\dots)$ should be closed 1-forms
in the functional space $\varphi(x)$. In the weakly-nonlocal
symplectic structures given by the integrable systems they usually
appear as the Euler-Lagrange derivatives of local Hamiltonian 
functionals of the corresponding hierarchy (see below).

\vspace{0.5cm}

\section{Weakly nonlocal Poisson and Symplectic Structures
and famous Integrable Systems.}
\setcounter{equation}{0}

\vspace{0.3cm}

 Let us mention here that  in the late 70s
many people considered an infinite series of the
the nonlocal Poisson Brackets  for the
famous integrable systems like KdV, NLS, etc., on the spaces
of rapidly decreasing functions. They obtained
these brackets following the Lenard-Magri scheme starting with 
the initial pair of local brackets for KdV (for the NLS 
system the second bracket is already nonlocal indeed, 
it is weakly nonlocal in our sense).
For the KdV - equation:
$\varphi_{t} = 6 \varphi \varphi_{x} - \varphi_{xxx} $
we have
$J_{0} = \partial $
(Gardner - Zakharov - Faddeev bracket)
and
$J_{1} = - \partial^{3}
+ 2 ( \varphi \partial + \partial \varphi ) $
(Lenard-Magri bracket)

 The recursion operator
$R = - \partial^{2} + 4 \varphi + 2 \varphi_{x}
\partial^{-1} $ such that $RJ_0=J_1$,
generates the next bracket with the Hamiltonian operator $R^2J_0=J_2$:

$$J_{2} = \partial^{5} - 8 \varphi \partial^{3}
- 12 \varphi_{x} \partial^{2} - 8 \varphi_{xx} \partial
+ 16 \varphi^{2} \partial - 2 \varphi_{xxx} +
16 \varphi \varphi_{x} - 4 \varphi_{x} \partial^{-1}
\varphi_{x} $$

 In  work ~\cite{enrubor} all nonlocal parts of higher PBs
 were calculated. These authors never defined any specific class
 of Poisson Brackets, but using their results we  are easily coming
 to the statement:

 {\it All  higher brackets for 
KdV given by the formula 
$J_{n} = R^{n} J_{0}$, $n \geq 0$ are weakly nonlocal.
The corresponding flows
$S^{i}_{(k)} (\varphi,\varphi_{x},\dots)$ are exactly the Higher
 KdV systems and the exact formula for $J^{n}$ can be written
in the form:

$$J^{n} = (local \,\, part) - \sum_{k=1}^{n-1}
S_{(k)}(\varphi,\varphi_{x},\dots) \partial^{-1}
S_{(n-k-1)}(\varphi,\varphi_{x},\dots)$$
where $S_{(1)}(\varphi,\varphi_{x},\dots) = 2\varphi_{x}$
and 

$$S_{(k)}(\varphi,\varphi_{x},\dots) \equiv R
S_{(k-1)}(\varphi,\varphi_{x},\dots)$$
are higher KdV flows.}

 The similar weakly-nonlocal expression for the positive powers 
of the recursion operator for KdV was also considered in
~\cite{enrubor}. Let us represent here the corresponding result

$$R^{n} = (local \,\, part) + \sum_{k=1}^{n}
S_{(k)}(\varphi,\varphi_{x},\dots) \partial^{-1}
{\delta H_{(n-k)} \over \delta \varphi(x)} \,\,\, , \,\,
n\geq 0$$
where $S_{(k)} = \partial_{x} \delta H_{(k)}/\delta\varphi(x)$,
$H_{(0)} = \int \varphi dx$ and

$${\delta H_{(k)} \over \delta \varphi(x)} \equiv 
{\delta H_{(k-1)} \over \delta \varphi(x)} R $$
are Euler-Lagrange derivatives of higher Hamiltonian
functions for KdV hierarchy. Let us mention also that in our
notations $R$ acts from the left on the vectors and from the right
on the 1-forms in the functional space.

 Using these results we prove here the following

 {\bf Proposition.}

{\it All  symplectic structures for KdV
$\Omega_{-n} = (J_{-n})^{-1}$ are weakly nonlocal for 
$n \geq 0$. The 1-forms $Q_{(k)}(\varphi,\varphi_{x},\dots)$
in the nonlocal part are the Euler-Lagrange derivatives of
higher Hamiltonian functions $H_{(n)}$ and the formula for
$\Omega_{-n}$ can be written as

$$\Omega_{-n} = (local \,\, part) + \sum_{k=0}^{n}
{\delta H_{(k)} \over \delta \varphi(x)} \partial^{-1}
{\delta H_{(n-k)} \over \delta \varphi(x)}$$

}

 Proof. We have $\Omega_{0} = \partial^{-1}$ and by the
definition:

$$\Omega_{-n} = \partial^{-1} R^{n} = 
\partial^{-1} (local \,\, part) + 
\sum_{k=1}^{n} \partial^{-1} S_{(k)}(\varphi,\varphi_{x},\dots)
\partial^{-1} {\delta H_{(n-k)} \over \delta \varphi(x)} =$$

$$= \partial^{-1} (local \,\, part) +
\sum_{k=1}^{n} \partial^{-1} 
\left({\delta H_{(k)} \over \delta \varphi(x)}\right)_{x}
\partial^{-1} {\delta H_{(n-k)} \over \delta \varphi(x)} =$$

$$= \partial^{-1} (local \,\, part) +
\sum_{k=1}^{n} \partial^{-1} \left( \partial
{\delta H_{(k)} \over \delta \varphi(x)} -
{\delta H_{(k)} \over \delta \varphi(x)} \partial \right)
\partial^{-1} {\delta H_{(n-k)} \over \delta \varphi(x)} =$$

$$= {\delta H_{(0)} \over \delta \varphi(x)} \partial^{-1}
{\delta H_{(0)} \over \delta \varphi(x)} (local \,\, part) +$$

$$+ \sum_{k=1}^{n} {\delta H_{(k)} \over \delta \varphi(x)}
\partial^{-1} {\delta H_{(n-k)} \over \delta \varphi(x)} -
\sum_{k=1}^{n} {\delta H_{(0)} \over \delta \varphi(x)}
\partial^{-1} {\delta H_{(k)} \over \delta \varphi(x)}
{\delta H_{(n-k)} \over \delta \varphi(x)} $$

 We can also write that

$${\delta H_{(0)} \over \delta \varphi(x)} (local \,\, part) =
\zeta(x) + \partial (local \,\, part)^{\prime} $$
where $\zeta(x)$ is a result of the action of the local part of
$R^{n}$ on $\delta H_{(0)}/\delta\varphi(x)$ (from the right)
and $(local \,\, part)^{\prime}$ is a differential operator.
So we have

$$\Omega_{-n} = (local \,\, part)^{\prime} + 
\sum_{k=1}^{n} {\delta H_{(k)} \over \delta \varphi(x)}
\partial^{-1} {\delta H_{(n-k)} \over \delta \varphi(x)} +$$

$$+ {\delta H_{(0)} \over \delta \varphi(x)} \partial^{-1}
\left[\zeta(x) - \sum_{k=1}^{n} 
{\delta H_{(k)} \over \delta \varphi(x)}
{\delta H_{(n-k)} \over \delta \varphi(x)} \right] $$
where the expression in the brackets is equal to

$${\delta H_{(0)} \over \delta \varphi(x)} R^{n} \equiv
{\delta H_{(n)} \over \delta \varphi(x)} $$

 So we obtain the statement of the Proposition.

\vspace{0.5cm}

 It seems that this property is very general for the brackets given
by the recursion operators for integrable systems. We prove
here the similar fact for  the NLS-equation:

$i \psi_{t} = - \psi_{xx} + 2 \kappa
|\psi|^{2} \psi $
where $\psi$ is a complex function. We have

\begin{equation}
\label{0nlsbr}
\{\psi(x), {\bar \psi}(y)\}_{0} =
i \delta (x-y) 
\end{equation}
This is the first Hamiltonian structure. It  corresponds 
to the operator

$$J_{0} = \left( \begin{array}{cc}
0 & i \cr -i & 0 \end{array} \right) $$
There is an infinite number of Hamiltonian structures
connected with this one through the recursion operator 
(~\cite{fadtah}, p. 218). The bracket 
$\{\dots,\dots\}_{1}$ corresponds to the Hamiltonian
operator

$$J_{1} = \left( \begin{array}{cc}
0 & \partial \cr \partial & 0 \end{array} \right) -
2 \kappa \left( \begin{array}{cc}
- \psi \partial^{-1} \psi & \psi \partial^{-1} {\bar \psi} \cr
{\bar \psi} \partial^{-1} \psi & 
- {\bar \psi} \partial^{-1} {\bar \psi} \end{array} \right) $$
which has the form (\ref{jform}) with only 
one flow in the nonlocal part

\begin{equation}
\label{nlstr}
\left( \begin{array}{c} \psi \cr {\bar \psi} \end{array} 
\right)_{t} =
\sqrt{2\kappa} \left( \begin{array}{c} i \psi \cr
-i {\bar \psi} \end{array} \right)
\end{equation}

\vspace{0.5cm}

{\bf Proposition.} {\it All the brackets 
$\{\dots,\dots\}_{n}$, $n \geq 0$ given by the recursion
$R^{n} J_{0}$ have the form (\ref{jform})
with the flows from NLS-hierarchy in the nonlocal parts.
All the symplectic structures 
$\Omega_{-n} = (J_{-n})^{-1} = \Omega_{0} R^{n}$ have the form
(\ref{wnsymp}) where the forms 
$Q_{(k)i}(\psi,{\bar \psi},\dots)$ are the Euler-Lagrange 
derivatives of the higher Hamiltonian functionals of the
NLS-hierarchy. The corresponding formulas for $J_{n}$,
$\Omega_{-n}$ and $R^{n}$ $(n\geq 0)$ can be written as

$$J_{n} = (local \,\, part) - \sum_{k=1}^{n} 
S_{(k-1)}(\psi,{\bar \psi},\dots) \partial^{-1}
S_{(n-k)}(\psi,{\bar \psi},\dots) $$

$$R^{n} = (local \,\, part) + \sum_{k=1}^{n}
S_{(k-1)}(\psi,{\bar \psi},\dots) \partial^{-1}
{\delta H_{(n-k)} \over \delta(\psi,{\bar \psi})(x)} $$

$$\Omega_{-n} = (local \,\, part) + \sum_{k=1}^{n}
{\delta H_{(k-1)} \over \delta(\psi,{\bar \psi})(x)}
\partial^{-1}
{\delta H_{(n-k)} \over \delta(\psi,{\bar \psi})(x)} $$
where 

$$S_{(k)} \equiv J_{0} 
{\delta H_{(k)} \over \delta(\psi,{\bar \psi})(x)} \,\, ,
\,\,\, H_{(0)} = \sqrt{2\kappa} \int \psi {\bar \psi} dx \,\, , 
\,\,\, and \,\,\, 
{\delta H_{(k)} \over \delta(\psi,{\bar \psi})(x)} =
R {\delta H_{(k-1)} \over \delta(\psi,{\bar \psi})(x)}$$
for any $k \geq 1$.

}

Proof.
We use the induction. The recursion operator
$R = J_{1} J_{0}^{-1}$ takes here the form:

$$R = \left( \begin{array}{cc}
- i \partial & 0 \cr 0 & i \partial \end{array} \right) +
2 \kappa \left( \begin{array}{cc}
i \psi \partial^{-1} {\bar \psi} & 
i \psi \partial^{-1} \psi \cr 
- i {\bar \psi} \partial^{-1} {\bar \psi} &
- i {\bar \psi} \partial^{-1} \psi \end{array} \right) $$

 Suppose now that $J_{n}$ has the required form:

\begin{equation}
\label{cfjn}
J_{n} = \sum_{k\geq0} B_{(n)k}
(\psi, {\bar \psi}, \dots) \partial^{k} -
\sum_{k=1}^{n} J_{0}
{\delta H_{(k-1)} \over \delta (\psi, {\bar \psi})(x)}
\partial^{-1} J_{0} 
{\delta H_{(n-k)} \over \delta (\psi, {\bar \psi})(x)}
\end{equation}
where $B_{(n)k}$ are $2 \times 2$ - matrices and
$H_{(k)}$ are the higher NLS Hamiltonians by
the induction assumption. 
We note that $R$ can be written
in the form:

$$R = \left( \begin{array}{cc}
- i \partial & 0 \cr 0 & i \partial \end{array} \right) +
\sqrt{2\kappa} \left( \begin{array}{cc}
i \psi \partial^{-1} \delta N/\delta \psi(x) &
i \psi \partial^{-1} \delta N/\delta {\bar \psi}(x) 
\cr
- i {\bar \psi} \partial^{-1} 
\delta N/\delta \psi(x) &
- i {\bar \psi} \partial^{-1} 
\delta N/\delta {\bar \psi}(x) \end{array} \right) $$
Here $N = \sqrt{2\kappa} \int \psi(x) {\bar \psi}(x) dx$ 
is the first Hamiltonian  in the NLS hierarchy, generating 
the flow (\ref{nlstr}) with respect to the 
bracket (\ref{0nlsbr}). Since it commutes with any 
$H_{(k)}$ with respect to the bracket (\ref{0nlsbr}), 
we have

$$\left({\delta N \over \delta \psi(x)} \,\,\,
{\delta N \over \delta {\bar \psi}(x)} \right) J_{0}
\left( {\delta H_{(k-1)} \over \delta \psi(x)} \,\,\,
{\delta H_{(k-1)} \over \delta {\bar \psi}(x)} \right)^{t}
\equiv \left( Q_{(k-1)} \right)_{x} = 
\partial Q_{(k-1)} - Q_{(k-1)} \partial $$
for some functions $Q_{(k-1)} (\psi, {\bar \psi}, \dots)$.
We have

$$R J_{n} = (local \,\, terms) - \sum_{k=1}^{n}
\left( \begin{array}{cc}
- i \partial & 0 \cr 0 & i \partial 
\end{array} \right)
J_{0} {\delta H_{(k-1)} \over \delta (\psi, {\bar \psi})(x)}
\partial^{-1} J_{0}
{\delta H_{(n-k)} \over \delta (\psi, {\bar \psi})(x)} +$$

$$+ \sqrt{2\kappa} \left( \begin{array}{cc}
i \psi \partial^{-1} \delta N/\delta \psi(x) &
i \psi \partial^{-1} \delta N/\delta {\bar \psi}(x)
\cr
- i {\bar \psi} \partial^{-1} \delta N/\delta \psi(x) &
- i {\bar \psi} \partial^{-1} \delta N/\delta {\bar \psi}(x) 
\end{array} \right) \sum_{k\geq0}
B_{(n)k}(\psi, {\bar \psi}, \dots) \partial^{k} - $$

$$- \sqrt{2\kappa} \sum_{k=1}^{n} 
\left( \begin{array}{c} i \psi \cr - i {\bar \psi}
\end{array} \right)
\partial^{-1} \left(
\partial Q_{(k-1)} - Q_{(k-1)} \partial \right)
\partial^{-1} J_{0} 
{\delta H_{(n-k)} \over \delta (\psi, {\bar \psi})(x)} $$

 We can write the relations:

$$\left( \begin{array}{cc}
- i \partial & 0 \cr 0 & i \partial
\end{array} \right) J_{0} 
{\delta H_{(k-1)} \over \delta (\psi, {\bar \psi})(x)} =
\left( \begin{array}{cc}
- i & 0 \cr 0 & i 
\end{array} \right)\left[
\left( J_{0}
{\delta H_{(k-1)} \over \delta (\psi, {\bar \psi})(x)}
\right)_{x} +  J_{0}
{\delta H_{(k-1)} \over \delta (\psi, {\bar \psi})(x)}
\partial \right]$$
and

$$\left( {\delta N \over \delta \psi (x)}
{\delta N \over \delta {\bar \psi} (x)} \right)
\sum_{k\geq0} B_{(n)k}(\psi, {\bar \psi}, \dots) \partial^{k} =
\left( \zeta(x) \,\, {\bar \zeta}(x) \right) + \partial
(local \,\, terms) $$
where $(\zeta(x) \,\, {\bar \zeta}(x))$ is a result of the action 
of the local part of $J_{n}$ (from the right) on 
$(\delta N/\delta \psi (x) \,\, \delta N/\delta {\bar \psi} (x))$.
So the nonlocal part of $RJ_{n}$ has a form

$$- \sum_{k=1}^{n} \left[ \left( \begin{array}{cc}
- i & 0 \cr 0 & i
\end{array} \right)
\left( J_{0}
{\delta H_{(k-1)} \over \delta (\psi, {\bar \psi})(x)}
\right)_{x} +
\sqrt{2\kappa} 
\left( \begin{array}{c}
i \psi \cr - i {\bar \psi}
\end{array} \right)
Q_{(k-1)} \right] \partial^{-1} J_{0}
{\delta H_{(n-k)} \over \delta (\psi, {\bar \psi})(x)} $$

$$+ \sqrt{2\kappa}
\left( \begin{array}{c}
i \psi \cr - i {\bar \psi}
\end{array} \right)
\partial^{-1} \left[ \left(\zeta(x) \,\, {\bar \zeta}(x)\right) +
\sqrt{2\kappa} \sum_{k=1}^{n}
Q_{(k-1)} J_{0}
{\delta H_{(n-k)} \over \delta (\psi, {\bar \psi})(x)} \right] $$

 Using the relations:

$$\left( \begin{array}{cc}
- i & 0 \cr 0 & i
\end{array} \right)
\left( J_{0}
{\delta H_{(k-1)} \over \delta (\psi, {\bar \psi})(x)}  
\right)_{x} + \sqrt{2\kappa}
\left( \begin{array}{c}
i \psi \cr - i {\bar \psi}
\end{array} \right)
Q_{(k-1)} = R S_{(k-1)} = S_{(k)} $$
and

$$\left(\zeta(x) \,\, {\bar \zeta}(x)\right) +
\sum_{k=1}^{n}
Q_{(k-1)} J_{0}
{\delta H_{(n-k)} \over \delta (\psi, {\bar \psi})(x)} =
{\delta N \over \delta (\psi, {\bar \psi})(x)} J_{n} =
- S_{(n)} $$
we obtain the required formula for $J_{n+1}$.

 Now using the expressions $R^{n} = J_{n} \Omega_{0}$ and
$\Omega_{-n} = \Omega_{0} R^{n} = \Omega_{0} J_{n} \Omega_{0}$
where

$$\Omega_{0} = \left( \begin{array}{cc}
0 & i \cr -i & 0  
\end{array} \right) $$
it is easy to obtain the corresponding formulas for
$R^{n}$ and $\Omega_{-n}$ for $n \geq 1$.

{\hfill Proposition is proved.}\footnote{
It is worth to mention that in the variables
$r = \sqrt{\psi {\bar \psi}}$, 
$\theta = -i(\psi_{x}/\psi - {\bar \psi}_{x}/{\bar \psi})$
(i.e. $\psi = r \exp(i\int \theta dx)$) the NLS-equation
has in fact three ($J_{0}$, $J_{1}$, $J_{2}$) local Hamiltonian
structures. We don't know any place in the literature where
this fact was clearly mentioned.}

\section{ Poisson brackets of Hydrodynamic Type.}
\setcounter{equation}{0}

 The well-known Whitham averaging procedure 
(or the nonlinear WKB me\-thod) we consider for
 the local (PDE) evolution system only. It is based on the family of
 invariant tori (i.e. exact solutions quasiperiodic in $x,t$)
  dependent on some
 parameters $U$. We assume that a family of the local integrals is given
 such that the value of parameters $U$ can be chosen as the average values
 of the densities of  integrals along the invariant tori.
 For the Hamiltonian systems we assume that these integrals are commuting
 (we call it ''Liouville Property''), and we have an invariant, completely
 integrable finite-dimensional subsystem or family of them.

  This procedure leads to
 first order homogeneous quasilinear systems 
(Hydrodynamic type system) useful in many cases for  asymptotic studies:

$$U^{\nu}_{T} = V^{\nu}_{\mu}(U) U^{\mu}_{X} $$

As it was established in 1983 (see ~\cite{dn1}), 
they are Hamiltonian in the
so-called Hydrodynamic Type Poisson Brackets (HTPB) or Dubrovin-Novikov
(DN)--brackets in the local
case (i.e. the original system was Hamiltonian in the local PB)

\begin{equation}
\label{dubrnov}
\{ U^{\nu}(X), U^{\mu}(Y) \} = g^{\nu\mu}(U)
\delta^{\prime}(X-Y) + b^{\nu\mu}_{\lambda}(U) U^{\lambda}_{X}
\delta (X-Y)
\end{equation}
introduced in ~\cite{dn1} (see also ~\cite{dn2}, ~\cite{dn3})
for  more complete information) .
(There was a gap in the  general proof of the Jacoby Identity
for the "averaged" Poisson Bracket
 constructed
in 1983; this gap was finally fulfilled  in work 
~\cite{engam}).

 Let us  remind here the  general properties of the brackets
 (\ref{dubrnov}) - (see ~\cite{dn1}-~\cite{dn3}).

   Any Hydrodynamic type Hamiltonian $H=\int h(U(x))dx$ 
(i.e. its density
does not depend on derivatives) generates a H.T.System
with Hydrodynamic type (or DN) P.B..
 Consider the H.T. bracket (\ref{dubrnov}) such that 
$det \,\, g^{\nu\mu} \neq 0$. 
It follows from the Leibnitz property that
the first coefficient ($\delta$-prime term) transforms under the
"pointwise" change of coordinates
$u(w)$ as a Riemannian metric with upper indices, and
the second coefficient ($\delta$-term) transforms as  a set of
Christoffel symbols (=connection) with two upper indices. Indices
should be raised up with the same  metric.
 It is skew-symmetric
if and only if  the tensor $g^{\nu\mu}$ is symmetric
(i.e. defines a pseudo-Riemannian metric), and  the connection
$\Gamma^{\nu}_{\mu\lambda} = - g_{\mu\tau} b^{\tau\nu}_{\lambda}$
is compatible with this metric: 
$\nabla_{\lambda}g_{\mu\nu} \equiv 0$.
 The bracket (\ref{dubrnov}) satisfies the Jacobi Identity if
and only if  connection $\Gamma^{\nu}_{\mu\lambda}$ is
symmetric, and the metric is flat.  
Therefore  the signature of metric is
a complete local invariant under the change of coordinates $u(w)$.

 Three types of coordinates for this kind of
P.B. play an important role in applications:

 1.The Canonical (flat) coordinates $n^{\mu}$  where 
 the connection coefficients are equal to zero. The integrals 
$\int n^{\nu}(x)dx$ are Casimirs
for this P.B.. 

 2. The so-called "Physical Coordinates"  
$U^{\nu}$ obtained by the averaging of densities of  local 
commuting integrals. We say that the coordinates are 
"Liouville" or Physical for H.T.P.B.  if it has the form:

$$\{U^{\nu}(X), U^{\mu}(Y)\} = (\gamma^{\nu\mu}(U) +
\gamma^{\mu\nu}(U)) \delta^{\prime}(X-Y) + 
{\partial \gamma^{\nu\mu} \over \partial U^{\lambda}} 
U^{\lambda}_{X} \delta (X-Y)$$
for some functions $\gamma^{\nu\mu}(U)$.

 Any coordinates such that integrals of them 
define the commuting flows, are physical in that sense.

 In particular, any coordinates such that  PB is linear, are physical.

  The general local Poisson Brackets of any order
linearly dependent on the fields were studied in 
works ~\cite{GD1,GD2}. The especially interesting class of 
linear Hydrodynamic Type P.B. was studied in  works
~\cite{No82}-~\cite{Ba}.
 It leads to the beautiful
algebraic and differentially geometrical theory of the local 
translational invariant first order Lie algebras, Frobenius 
type algebras and their non-associative
analogs ("Novikov Algebras"), and super-analogs of that theory. 

 3. The diagonal  form where our metric and H.T.system both are 
diagonal: the coordinates are  orthogonal for 
the metric, and the velocity tensor $V^{\mu}_{\nu}(U)$ is diagonal 
(in  classical terminology the coordinates $U$ are the
"Riemann Invariants" for our system).
This form has the following fundamental property: all
diagonal H.T. Systems Hamiltonian in such PB, are Completely 
Integrable. It was conjectured by S.P.Novikov and proved by 
S.P. Tsarev in his PhD thesis in 1985 (see ~\cite{tsarev}).

   Let us point out that the beautiful Tsarev integration 
procedure  based on the Riemannian metric turns out to be
more broad than the class of H.T.Systems Hamiltonian 
corresponding  to the local H.T.P.B.. It integrates also the
 systems  called
"semihamiltonian". The Riemannian metric is nonflat in that 
case. Probably, {\bf all semihamiltonian systems are in fact 
Hamiltonian corresponding to some weakly nonlocal H.T.P.B. 
with (maybe) an infinite number of terms in the  nonlocal tail.}
Some investigation of this problem can be found in
~\cite{fer4}, ~\cite{bfer} but 
this problem is still open.

 The first weakly nonlocal fundamental PB of hydrodynamic 
type was found in ~\cite{mohfer1},
the more general class - in ~\cite{fer1}-~\cite{fer4}. 
The weakly nonlocal PB of the
types different from the hydrodynamic one,
never have been studied to our knowledge.
Let us describe  the Mokhov-Ferapontov fundamental 
weakly nonlocal Poisson
bracket of hydrodynamic type (MF-bracket,
see ~\cite{mohfer1}):

$$\{ U^{\nu}(X), U^{\mu}(Y) \} = g^{\nu\mu}(U)
\delta^{\prime}(X-Y) + b^{\nu\mu}_{\lambda}(U) U^{\lambda}_{X}
\delta (X-Y) + $$
\begin{equation}
\label{mokfer}
+ c U^{\nu}_{X} \nu (X-Y) U^{\mu}_{Y}
\end{equation}
and more general Ferapontov weakly nonlocal brackets  
of hydrodynamic type
(F-bracket ~\cite{fer1}):
$$\{ U^{\nu}(X), U^{\mu}(Y) \} = g^{\nu\mu}(U)
\delta^{\prime}(X-Y) + b^{\nu\mu}_{\lambda}(U) U^{\lambda}_{X}
\delta (X-Y) + $$
\begin{equation}
\label{fer}
+ \sum_{k=1}^{g} e_{k}  w^{\nu}_{(k)\lambda}(U)
U^{\lambda}_{X}\,\, \nu(X-Y)\,\, w^{\mu}_{(k)\delta}(U) 
U^{\delta}_{Y}
\end{equation} 
$e_{k} = \pm 1$.

 Consider the fundamental MF-bracket (\ref{mokfer}) with 
non-degenerate metric tensor $g^{\nu\mu}(U)$. 
It is skew-symmetric
if and only if the tensor $g^{\nu\mu}$ is symmetric,
and the connection

$$\Gamma^{\nu}_{\mu\lambda} = - g_{\mu\tau} b^{\tau\nu}_{\lambda}$$ 
is compatible with this metric:
$\nabla_{\lambda}g_{\mu\nu} \equiv 0$.

 The bracket (\ref{mokfer}) satisfies  Jacobi Identity
if and only if  its connection $\Gamma^{\nu}_{\mu\lambda}$
is symmetric   (i.e. torsion tensor is equal to zero)
and has constant curvature equal to $c$ , i.e.

\begin{equation}
\label{concurv}
R^{\nu\tau}_{\mu\lambda} = c 
\left(\delta^{\nu}_{\mu} \delta^{\tau}_{\lambda} -
\delta^{\tau}_{\mu} \delta^{\nu}_{\lambda} \right)
\end{equation}

\vspace{0.5cm}

 Consider the F-bracket (\ref{fer}) with 
$det \,\, g^{\nu\mu} \neq 0$.

 The bracket (\ref{fer}) is skew-symmetric
if and only if  tensor $g^{\nu\mu}$ is symmetric
 and  connection
is compatible with this metric as above for the local case.

 The bracket (\ref{fer}) satisfies  Jacobi Identity
if and only if its connection $\Gamma^{\nu}_{\mu\lambda}$
is symmetric, the metric $g^{\nu\mu}$ (with lower indices)
and  tensors
$w^{\nu}_{(k) \mu}$ satisfy the equations:

\begin{equation}
\label{metwein}
g_{\nu\tau} w^{\tau}_{(k) \mu} = g_{\mu\tau} w^{\tau}_{(k) \nu},
\,\,\,\,\, \nabla_{\nu}  w^{\mu}_{(k) \lambda} =
\nabla_{\lambda}  w^{\mu}_{(k) \nu} 
\end{equation}

\begin{equation}
\label{rtenw}
R^{\nu\tau}_{\mu\lambda} = \sum_{k=1}^{g} e_{k}
\left(w^{\nu}_{(k) \mu} w^{\tau}_{(k) \lambda} -
w^{\tau}_{(k) \mu} w^{\nu}_{(k) \lambda} \right) 
\end{equation}

 Moreover, this set  is commutative
$[w_{k}, w_{k^{\prime}}] = 0$.

 It was  pointed out by E.V.Ferapontov that the equations
written above are  the Gauss-Codazzi equations for the
submanifolds ${\cal M}^{N}$ with flat normal connection in
the Pseudo-Euclidean space $E^{N+g}$. Here $g_{\nu\mu}$ is the 
first quadratic form of ${\cal M}^{N}$, and $w_{(k)}$ are the 
Weingarten operators corresponding to the field of pairwise 
orthogonal unit normals ${\vec n}_{k}$, see
~\cite{fer1}-~\cite{fer4}. Moreover, it was proved by 
E.V.Ferapontov that these brackets can be obtained as a result of 
Dirac restriction of the local DN-bracket

$$\{N^{I}(X), N^{J}(Y)\} = \epsilon^{I} \delta^{IJ} 
\delta^{\prime}(X-Y), \,\,\, I, J = 1,\dots , N+g, \,\,\,
\epsilon^{I} = \pm 1 $$
in $E^{N+g}$ to the submanifold ${\cal M}^{N}$ (see
~\cite{fer2},~\cite{fer4}).
 
 Let us note that for the brackets
(\ref{fer}) corresponding to the submanifolds with flat
normal connection and  holonomic net of the lines of curvature,
the commutativity of the
flows $w^{\nu}_{(k)\mu}u^{\mu}_{X}$ was proved in
 ~\cite{fer1})  for this particular case.
 Concerning their Hamiltonian
structure, it was suggested in  work ~\cite{fer1}  to consider 
them as "being generated by formal Hamiltonian functions like
$H = \int 1 dx$". This statement  makes no sense in any
symplectic geometry where the Poisson Bracket is  well-defined.
Besides that, we demonstrate below the local Hamiltonians
 generating these flows.

\vspace{0.5cm}

 Let us introduce the  {\bf Physical} or {\bf Liouville} 
coordinates for the weakly nonlocal Poisson brackets of 
hydrodynamic type: this form is given  by the formulas

$$\{U^{\nu}(X),U^{\mu}(Y)\} =
\left(\gamma^{\nu\mu}(X) + \gamma^{\mu\nu}(X)
- \sum_{k=1}^{g} e_{k} f^{\nu}_{(k)} f^{\mu}_{(k)}
\right)
\delta^{\prime}(X-Y) + $$
\begin{equation}
\label{nphform}
+ \left(
{\partial \gamma^{\nu\mu} \over \partial U^{\lambda}}
U^{\lambda}_{X} - \sum_{k=1}^{g} e_{k} (f^{\nu}_{(k)})_{X}
f^{\mu}_{(k)}\right) \delta (X-Y) +
\sum_{k=1}^{g} e_{k} (f^{\nu}_{(k)})_{X} \nu(X-Y)
(f^{\mu}_{(k)})_{Y}
\end{equation}
for some functions $\gamma^{\nu\mu}(U)$ and $f^{\nu}_{(k)}(U)$.
Like in the local case, we have the following property:

\vspace{0.5cm}

 {\bf Lemma 1.} {\it 
The bracket (\ref{fer}) of F-type (i.e. weakly nonlocal
H.T.P.B.) has  Physical form
in the coordinates $U^{\mu}$ if and only if the
integrals $J^{\nu} = \int U^{\nu}(X) dX$
generate the set of local commuting
flows according to the bracket (\ref{fer}).}

\vspace{0.5cm}

 Proof. Suppose all functionals $J^{\nu}$ satisfy the
 conditions above. Then we have
for any $\nu$ and $k$

$$w^{\nu}_{(k)\tau} (U) U^{\tau}_{X} \equiv
f^{\nu}_{(k)X} $$
for some functions $f^{\nu}_{(k)}(U)$ as it follows from the 
condition of locality of the flow corresponding to $J^{\nu}$. 
 From the commutativity of the set of functionals $\{J^{\nu}\}$ we 
then have

$$b^{\nu\mu}_{\lambda} U^{\lambda}_{X} +
\sum_{k\geq 0} e_{k} (f^{\nu}_{(k)})_{X}f^{\mu}_{(k)}
\equiv {d \over dx} \gamma^{\nu\mu} $$
for some $\gamma^{\nu\mu}(U)$. Again, 
from the skew-symmetry property of the bracket we have

$${d \over dx} g^{\nu\mu} \equiv
b^{\nu\mu}_{\lambda} U^{\lambda}_{X} +
b^{\mu\nu}_{\lambda} U^{\lambda}_{X} =
{d \over dx} \left(\gamma^{\nu\mu} + \gamma^{\mu\nu} -
\sum_{k\geq 0} e_{k} f^{\nu}_{(k)} f^{\mu}_{(k)} \right) $$
and it's clear that we can choose $\gamma^{\nu\mu}$ in such
a way that

$$g^{\nu\mu} = \gamma^{\nu\mu} + \gamma^{\mu\nu} -
\sum_{k\geq 0} e_{k} f^{\nu}_{(k)} f^{\mu}_{(k)} $$

It is easy to obtain from (\ref{nphform}) that functionals
$J^{\nu}$
generate the local flows according to this bracket and commute
with each other.

{\hfill Lemma is proved.}

\vspace{0.5cm}

 Let us also point out here that the first non-local bracket for
the averaged NLS-equation was constructed by M.V.Pavlov
(~\cite{pavlov}) from a nice differential-geometrical
consideration. In ~\cite{alekspav} and ~\cite{alekseev}
 the construction of non-local brackets for the averaged
KdV equation  using its local bi-Hamiltonian 
structure was considered.

\section{\centerline{Canonical forms and Casimirs}
\centerline{for Fundamental Poisson bracket.}}
\setcounter{equation}{0}

 The structure of the MF 
fundamental bracket written in the densities of Casimirs
was first found by M.V. Pavlov
in  Theorem 6 of  work ~\cite{pavlov2}   who never published
any proof later. As far as we know, he never paid any attention to 
the dependence on the
boundary conditions and linear Poisson 
$\lambda$-pencils in this structure.

 {\bf Definition.} {\it
Let   MF type bracket (\ref{mokfer}) be given
with non-degenerate metric tensor $g^{\nu\mu}(U)$. We say that 
it is written in the canonical form if:

$$\{n^{\nu}(X), n^{\mu}(Y)\} = $$
\begin{equation}
\label{mfercan}
= \left( \epsilon^{\nu} 
\delta^{\nu\mu} - c n^{\nu} n^{\mu}\right)
\delta^{\prime}(X-Y) - c n^{\nu}_{X} n^{\mu}
\delta (X-Y) + 
c n^{\nu}_{X} \nu (X-Y) n^{\mu}_{Y} 
\end{equation}
where $\epsilon^{\nu}$ are equal to  $\pm 1$, and the term

$\epsilon^{\nu} \delta^{\nu\mu}$ has the same signature
as metric tensor $g^{\nu\mu}(U)$.
}

 Let us formulate here the following Theorem:

\vspace{0.5cm}

{\bf Theorem 1.} {\it
Every fundamental MF-bracket (\ref{mokfer}) 
with non-degenerate metric tensor $g^{\nu\mu}(U)$
of the constant curvature $c$ can be written
locally in the canonical form (\ref{mfercan}) 
for any point $U_{0}$ after some
change of coordinates 
$n^{\nu} = n^{\nu}(U^{1},\dots,U^{N})$, such that 
$n^{\nu}(U_{0}) \equiv 0$.

 The bracket (\ref{mfercan})
represents  the {\bf Linear Poisson pencil}
given by the
compatible Poisson brackets

$$\{n^{\nu}(X), n^{\mu}(Y)\}_{0} = \epsilon^{\nu}
\delta^{\nu\mu} \delta^{\prime}(X-Y) $$
and

$$\{n^{\nu}(X), n^{\mu}(Y)\}_{1} = $$
$$ = - n^{\nu}(X) n^{\mu}(X)
\delta^{\prime}(X-Y) - n^{\nu}_{X} n^{\mu}(X) \delta (X-Y)
+ n^{\nu}_{X} \nu (X-Y) n^{\mu}_{Y} $$
where  curvature $c$ is the parameter of a pencil.
}

\vspace{0.5cm}

 We postpone the proof of Theorem 1 for the general case of
Ferapontov brackets (see Theorem 3). Here we just check that the
expression (\ref{mfercan}) really defines the MF-bracket 
with constant curvature $c$. Here we have:

$$g^{\nu\mu} = \epsilon^{\nu}
\delta^{\nu\mu} - c n^{\nu} n^{\mu} $$
where $\epsilon^{\nu}= \pm 1$, and

$$b^{\nu\mu}_{\lambda} = - c \delta^{\nu}_{\lambda} n^{\mu} $$

 The symmetry of $g^{\nu\mu}$ is clear;  the compatibility
of the connection with  metric can be written in upper
indices  as an equation

$${\partial g^{\nu\mu} \over \partial n^{\lambda}} =
b^{\nu\mu}_{\lambda} + b^{\mu\nu}_{\lambda} $$
(see ~\cite{dn2}), which is also clear in our case.
So we should check  symmetry of the connection and its 
 curvature. The symmetry of connection in
upper indices is equivalent to the equation:

$$b^{\nu\mu}_{\alpha} g^{\alpha\lambda} -
b^{\lambda\mu} g^{\alpha\nu} = 0 $$
which in our case is

$$- c \delta^{\nu}_{\alpha} n^{\mu} \left(
\epsilon^{\alpha} \delta^{\alpha\lambda} -
c n^{\alpha} n^{\lambda} \right) + 
c \delta^{\lambda}_{\alpha} n^{\mu} \left(
\epsilon^{\alpha} \delta^{\alpha\nu} - 
c n^{\alpha} n^{\nu} \right) = $$

$$= - c n^{\mu} \epsilon^{\nu} \delta^{\nu\lambda} +
c^{2} n^{\mu} n^{\nu} n^{\lambda} +
c n^{\mu}\epsilon^{\lambda} \delta^{\lambda\nu} -
c^{2} n^{\mu} n^{\lambda} n^{\nu} \equiv 0 $$

 For the curvature we use the formula 

$$- g^{\nu\alpha} g^{\mu\beta} R^{\tau}_{\beta\lambda\alpha}
= \left( b^{\mu\tau}_{\lambda , \alpha} -
b^{\mu\tau}_{\alpha , \lambda} \right) g^{\alpha\nu} +
b^{\nu\mu}_{\alpha}b^{\alpha\tau}_{\lambda} -
b^{\nu\tau}_{\alpha}b^{\alpha\mu}_{\lambda} = $$

$$= \left( - c \delta^{\mu}_{\lambda} \delta^{\tau}_{\alpha} +
c \delta^{\mu}_{\alpha} \delta^{\tau}_{\lambda} \right)
g^{\alpha\nu} + c^{2} \delta^{\nu}_{\alpha} n^{\mu}
\delta^{\alpha}_{\nu} n^{\tau} - c^{2} 
\delta^{\nu}_{\alpha} n^{\tau} \delta^{\alpha}_{\lambda}
n^{\mu} \equiv $$

$$ \equiv - c \left( \delta^{\mu}_{\lambda}
\delta^{\tau}_{\alpha} - \delta^{\tau}_{\lambda}
\delta^{\mu}_{\alpha} \right) g^{\alpha\nu} $$
so that finally we have

$$ R^{\mu\tau}_{\lambda\alpha} = g^{\mu\beta}
R^{\tau}_{\beta\lambda\alpha} = c \left(
\delta^{\mu}_{\lambda} 
\delta^{\tau}_{\alpha} - \delta^{\tau}_{\lambda}
\delta^{\mu}_{\alpha} \right) $$
which corresponds to the constant curvature $c$.

\vspace{0.5cm}

{\bf Remark}. The  metric
$g^{\nu\mu}$ in (\ref{mfercan}) is not well-defined for all 
values of $n^{\nu}$ but only for some small enough domain near 
the zero point.
So our theorem claims in fact that for the space of constant
curvature ${\cal M}^{N}$ (and given signature of the metric)
there exists a standard Canonical form for the corresponding  
bracket (\ref{mokfer})
on the small neighborhood of the constant point in the  
"loop space", i.e  space of mappings
$R^{1} \rightarrow {\cal M}^{N}$, such that
$n(X) \rightarrow y$ at $X \rightarrow \pm \infty$
for some point $y \in {\cal M}^{N}$ .
We shall clarify in the next work the canonical forms 
globally.

 In  work ~\cite{pavlov2} an implicit expression
for the density of the momentum functional for the bracket
(\ref{mfercan}) in terms of annihilators was also derived.
\footnote{As it will be clear from our later considerations
we can not actually speak about the Casimirs
and momentum functional until we fix the boundary conditions at
infinity. As we shall show in the general case it's better
to speak about the invariant set of $N+g$ canonical functionals
playing the role of annihilators or some 
"canonical Hamiltonian functions"
depending on the boundary conditions. We postpone these 
considerations to the case of the general Ferapontov bracket
and assume here the rapidly decreasing boundary conditions
at infinity.} We formulate here this result in the explicit form.

\vspace{0.5cm}

{\bf Lemma 2.} {\it
On the space of rapidly decreasing 
functions $n^{\nu}(X)$ the functionals

\begin{equation}
\label{ann}
N^{\nu} = \int n^{\nu} dX
\end{equation}
are the annihilators of the bracket (\ref{mfercan}). The
functional

\begin{equation}
\label{moment}
P = {1 \over c}  \int \left(1 - 
\sqrt{|1 - c \sum_{\nu = 1}^{N} \epsilon^{\nu}
n^{\nu}(X) n^{\nu}(X)|} \right)  dX
\end{equation}
is the momentum  generating  shifts along the 
coordinate $X$.
}

\vspace{0.5cm}

 This lemma follows from a simple calculation. 
We just point out here that the constant in the density 
${\cal P}(n)$ of the functional $P$ is chosen in such a way
that ${\cal P}(X) \rightarrow 0$ at $X \rightarrow \pm \infty$
on the space of rapidly decreasing at 
$X \rightarrow \pm \infty$ functions $n^{\nu}(X)$.
The expression (\ref{moment}) becomes the ordinary momentum
operator

$${1 \over 2} \int \sum_{\nu = 1}^{N} \epsilon^{\nu}
n^{\nu}(X) n^{\nu}(X) dX $$
for the corresponding DN bracket
as $c \rightarrow 0$.
Our definition of the operator ${\partial}^{-1}$ leads to the 
identity:

$$\sum_{\mu} {\partial}^{-1} n^{\mu}_{X} 
{\partial {\cal P}(n) \over \partial n^{\mu}} \equiv
{\cal P}(n) $$

 It is interesting  that the functional $P$ is uniquely
defined only on the part of the phase space where

$$\sum_{\nu = 1}^{N} \epsilon^{\nu}
n^{\nu}(X) n^{\nu}(X) < 1/c $$
for any $X$. If it is not so, we have nonsmooth functions and the 
problem of choosing the sign. Let us postpone these questions.

  The bracket (\ref{mfercan}) is written also in the
Physical form with 
$\gamma^{\nu\mu}(n) = 1/2 \,\, \epsilon^{\nu} \delta^{\nu\mu}$,
$e = sgn \,\, c$ and $f^{\nu}(n) = \sqrt{|c|} \,\, n^{\nu}$.

\vspace{0.5cm}

{\bf Lemma 3.} {\it
Suppose  we have a Poisson bracket (\ref{nphform}) such that

$$\gamma^{\nu\mu}(n) = A^{\nu\mu} + B^{\nu\mu}_{\lambda} 
n^{\lambda}$$ 
for some constants $A^{\nu\mu}$ and $B^{\nu\mu}_{\lambda}$,
$e = sgn \,\, c$ and $f^{\nu}(n) = \sqrt{|c|} \,\, n^{\nu}$.
Then  
$B^{\nu\mu}_{\lambda} \equiv \delta^{\nu}_{\lambda} b^{\mu}$
for some constants $b^{\mu}$; therefore the linear part can 
be removed from the bracket  by the simple shifts of coordinates
$n^{\nu} = {\tilde n}^{\nu} + b^{\nu}/c$.
}

\vspace{0.5cm}

Proof. We have in our case

$$g^{\nu\mu}(n) = A^{\nu\mu} + A^{\mu\nu} + 
B^{\nu\mu}_{\lambda} n^{\lambda} + 
B^{\mu\nu}_{\lambda} n^{\lambda} - c n^{\nu} n^{\mu} $$

$$b^{\nu\mu}_{\lambda} = B^{\nu\mu}_{\lambda} - 
c \delta^{\nu}_{\lambda} n^{\mu} $$

 So from the symmetry of the connection 

$$b^{\nu\mu}_{\alpha} g^{\alpha\lambda} - 
b^{\lambda\mu}_{\alpha} g^{\alpha\nu} = 0$$
we can obtain for the quadratic  part in the variables $n$:

$$B^{\nu\mu}_{\alpha} n^{\alpha} n^{\lambda} - 
B^{\lambda\mu}_{\beta} n^{\beta} n^{\nu} \equiv 0$$
for $\nu \neq \lambda$. So for $\alpha \neq \nu$ or
$\beta \neq \lambda$ we obtain 

$$B^{\nu\mu}_{\alpha} \equiv 0 \,\,\, , \,\,\,\,\,
B^{\lambda\mu}_{\beta} \equiv 0$$
and for $\alpha = \nu$ and $\beta = \lambda$

$$B^{\nu\mu}_{\nu} = B^{\lambda\mu}_{\lambda}$$
for any $\nu \neq \lambda$. So we have
$B^{\nu\mu}_{\lambda} \equiv \delta^{\nu}_{\lambda} b^{\mu}$ 
for some $b^{\mu}$.

{\hfill Lemma is proved.}

\vspace{0.5cm}

 It is clear also that any change of the term 
$\epsilon^{\nu} \delta^{\nu\mu}$ by a constant symmetric
matrix $A^{\nu\mu}$ leads  to the Poisson pencil;
however, in the degenerate case we do not claim that any bracket
(\ref{mokfer}) of the MF type can be presented in 
that form after some coordinate transformation.

 The  MF-brackets can be obtained as the
averaged ones according to the recent works
~\cite{malnloc1}-~\cite{malnloc2} from the original 
weakly nonlocal fundamental 
brackets with terms like

$$\varphi^{i}_{x} \nu (x-y) \varphi^{j}_{y}$$
in the non-local part.

\section{\centerline{General F-brackets.} 
\centerline{Riemannian Geometry.}}
\setcounter{equation}{0}

 Let us  consider more general F-brackets.

\vspace{0.5cm}

{\bf Definition.} {\it
We  say that the F-bracket is 
written in the Canonical form if
$$\{n^{\nu}(X), n^{\mu}(Y)\} = \left( \epsilon^{\nu}
\delta^{\nu\mu} - \sum_{k=0}^{g} e_{k} f^{\nu}_{(k)}(n) 
f^{\mu}_{(k)}(n) \right)
\delta^{\prime}(X-Y) - $$
$$- \sum_{k=0}^{g} e_{k}
\left( f ^{\nu}_{(k)}(n) \right)_{X} f^{\mu}_{(k)}(n)
\delta (X-Y) + $$
\begin{equation}
\label{fercan}
+\sum_{k=0}^{g} e_{k} 
\left( f ^{\nu}_{(k)}(n) \right)_{X} \nu (X-Y) 
\left( f ^{\mu}_{(k)}(n) \right)_{Y}
\end{equation}
with non-degenerate metric and some
 functions $f^{\nu}_{(k)}(n)$ such that
$f^{\nu}_{(k)}(0) \equiv 0$,   $e_{k} = \pm 1$.
}

\vspace{0.5cm}

{\bf Theorem 2.} {\it
The expression (\ref{fercan}) with non-degenerate metric
and linearly independent set of functions 
$f^{\nu}_{(k)}(n)$ defines a Poisson bracket if and
only if:

1) The flows 
$$n^{\nu}_{T_{k}} = \left( f ^{\nu}_{(k)}(n) \right)_{X}$$
commute with each other, i.e.
$d/dT_{k^{\prime}} (f ^{\nu}_{(k)}(n)) = d/dT_{k} (f
^{\nu}_{(k^{\prime})}(n))$
or

$${\partial f ^{\nu}_{(k)} \over \partial n^{\lambda} }
{\partial f ^{\lambda}_{(k^{\prime})} \over \partial n^{\mu} }
- {\partial f ^{\nu}_{(k^{\prime})} \over \partial n^{\lambda} }
{\partial f ^{\lambda}_{(k)} \over \partial n^{\mu} } 
\equiv 0 $$
for any $k$, $k^{\prime}$.

 The functions $f ^{\nu}_{(k)}(n)$ are such that 2) and 3) 
are true:

2)
$${\partial f ^{\nu}_{(k)} \over \partial n^{\lambda} }
\left(\epsilon^{\lambda} \delta^{\lambda\mu} - 
\sum_{s=0}^{g} e_{s}
f^{\lambda}_{(s)} f^{\mu}_{(s)} \right) = 
{\partial f ^{\mu}_{(k)} \over \partial n^{\lambda} }
\left(\epsilon^{\lambda} \delta^{\lambda\nu} -
\sum_{s=0}^{g} e_{s}
f^{\lambda}_{(s)} f^{\nu}_{(s)} \right) $$

3) 
$$\left( \epsilon^{\lambda} \delta^{\lambda\alpha} -
\sum_{s=0}^{g} e_{s} f^{\lambda}_{(s)}f^{\alpha}_{(s)}
\right) \left( \sum_{n=0}^{g} e_{n}
{\partial f ^{\mu}_{(n)} \over \partial n^{\beta} }
f ^{\nu}_{(n)} \right)
{\partial f ^{\beta}_{(k)} \over \partial n^{\alpha} } = $$

$$= \left( \epsilon^{\mu} \delta^{\mu\alpha} -
\sum_{s=0}^{g} e_{s} f^{\mu}_{(s)}f^{\alpha}_{(s)}
\right) \left( \sum_{n=0}^{g} e_{n}
{\partial f ^{\lambda}_{(n)} \over \partial n^{\beta} }
f ^{\nu}_{(n)} \right)
{\partial f ^{\beta}_{(k)} \over \partial n^{\alpha} } $$
}

\vspace{0.5cm}

Proof. As was shown in ~\cite{malnloc2},
the commutativity of the flows $(f ^{\nu}_{(k)})_{X}$
is the necessary requirement under the conditions 
of the Theorem. 

The condition $(2)$ is exactly the identity

$$g_{\nu\alpha} w^{\alpha}_{(k)\mu} =
g_{\mu\alpha} w^{\alpha}_{(k)\nu}$$
written in the upper indices for our case.

 By  direct calculation (like in Theorem 1) it's not
hard to show that in our case the conditions

$$R^{\nu\tau}_{\mu\lambda} = \sum_{k=1}^{g} e_{k}
\left(w^{\nu}_{(k) \mu} w^{\tau}_{(k) \lambda} -
w^{\tau}_{(k) \mu} w^{\nu}_{(k) \lambda} \right) $$
and the symmetry of connection are  corollaries of
$(1)$ and $(2)$, respectively.

 The condition $(3)$ is equivalent to the equality

$$\nabla_{\mu} w^{\nu}_{(k)\lambda} =
\nabla_{\lambda} w^{\nu}_{(k)\mu} $$

 So from  Ferapontov's results we obtain the statement 
of our Theorem.
 
Let us  mention also that the commutativity of affinors 
$[w_{(k)}, w_{(k^{\prime})}] = 0$ follows from the commutativity 
of the flows $w^{\nu}_{(k)\mu}n^{\mu}_{X}$.

{\hfill Theorem is proved.}

\vspace{0.5cm}

 Let the F-bracket (\ref{fer}) be given with
the non-degenerate tensor $g^{\nu\mu}(U)$. Fix some
point $U_{0} = (U_{0}^{1},\dots,U_{0}^{N})$. By
the local Canonical Form of F-bracket
corresponding to the point $U_{0}$ we call the bracket 
(\ref{fercan}) where $n^{\nu} = n^{\nu}(U)$,
$n^{\nu}(U_{0}) \equiv 0$; the term 
$\epsilon^{\nu} \delta^{\nu\mu}$ has the same
signature as $g^{\nu\mu}(U)$ and 
$f^{\nu}_{(k)}(U_{0}) \equiv 0$ at the point $U_{0}$.

\vspace{0.5cm}

{\bf Theorem 3.} {\it

I) Every F-bracket (\ref{fer}) 
with the non-degenerate metric 
tensor $g^{\nu\mu}(U)$ can be locally written in the Canonical
form (\ref{fercan}) after some
coordinate transformation $n^{\nu} = n^{\nu}(U)$. 
Moreover, for any given point $U_{0}$ it's possible to choose the
coordinates $n^{\nu}(U)$ in such a way that 
$n^{\nu}(U_{0}) \equiv 0$, $f^{\nu}_{(k)}(U_{0}) \equiv 0$. 

II) The integrals 

\begin{equation}
\label{ferann}
N^{\nu} = \int n^{\nu}(X) dX 
\end{equation}
are  annihilators of the bracket (\ref{fercan}) on the  
domain in the space of rapidly decreasing functions 
$n^{\nu}(X)$ bounded by the small enough constant;

III) The flows

$$n^{\nu}_{t_{k}} = {d \over dX} f^{\nu}_{(k)}(n) $$
are generated by the local Hamiltonians

$$H_{k} = \int h_{k}(n) dX $$
on the same phase space.  The functions
$n^{\nu}(U)$, $h_{k}(n(U))$ can be represented as  
linear combinations of coordinates $V^{I}$ in the
pseudo-Euclidean space $R^{N+g}$ for the local representation 
of our manifold as a submanifold $M^{N} \subset R^{N+g}$ with
flat normal connection.
}

\vspace{0.5cm}

 Proof. 

 As follows from the results of Ferapontov (~\cite{fer4})
the relations (\ref{metwein})-(\ref{rtenw}) are 
the Gauss-Codazzi equations; so 
every space with non-degenerate metric $g^{\nu\mu}(U)$ and the 
relations (\ref{metwein})-(\ref{rtenw}) for the 
operators $w^{\nu}_{(k)\mu}(U)$
can be represented as a submanifold ${\cal M}^{N}$ with 
flat normal connection in the pseudo-Euclidean space $R^{N+g}$ 
such that:

 the restriction of the constant metric on ${\cal M}^{N}$ is 
equal to $g^{\nu\mu}(U)$;

 the restriction of this  metric on the 
orthogonal subspace has at every point the 
signature corresponding
to the signature of the nonlocal part of the bracket (\ref{fer});

$w^{\nu}_{(k)\mu}(U)$ are the Weingarten operators
corresponding to $g$ parallel normal vector fields.  

Moreover, the bracket (\ref{fer})  can be obtained as a Dirac
restriction of the constant bracket

\begin{equation}
\label{brev}
\{V^{I}(X), V^{J}(Y)\} = e_{I} \delta^{IJ}
\delta^{\prime}(X-Y)
\end{equation}
 where $I,J = 1,\dots,N+g$,
$e_{I} = \pm 1$.

 Let us  point out also
that the Dirac restriction of (\ref{brev}) to any submanifold 
${\cal M}^{N}$ (not necessarily with flat normal connection) has 
the similar structure as (\ref{fer}) with $\partial^{-1}$ replaced
by $\nabla_{\perp}^{-1}$ where $\nabla_{\perp}$ is the normal 
connection corresponding to ${\cal M}^{N}$ 
(Ferapontov ~\cite{fer4}).

We can assume that the coordinates $V^{I}$ are such that:

1) $V^{I} \equiv 0$ at the point $U_{0}$ of submanifold
${\cal M}^{N}$;

2) The first $N$ coordinate lines are 
tangent to ${\cal M}^{N}$
at the point $U_{0}$;  the remaining $g$ coordinates are 
orthogonal to ${\cal M}^{N}$.

 It  follows that the first $N$ quantities $e_{I}$ in
this new system of coordinates  correspond to the signature
of the metric $g^{\nu\mu}(U)$, and the last $g$ coordinates 
correspond to the nonlocal part of (\ref{fer}).

 We take $n^{\nu}(U) = V^{\nu}$, $\nu = 1,\dots,N$ as the
local coordinates on the manifold ${\cal M}^{N}$ in the 
domain around of $U_{0}$
($n^{\nu}(U_{0}) \equiv 0$); for the last of the quantities
$V^{I}$ at the
points of ${\cal M}^{N}$ we have

$$V^{N+k} = v_{k}(n)$$
for some functions $v_{k}(n)$ such that $v_{k}(U_{0}) = 0$,
$\partial v_{k}/\partial n^{\nu} (U_{0}) = 0$.

 Make now the Dirac restriction of the bracket 
(\ref{brev}) on ${\cal M}^{N}$ in the coordinates $n^{\nu}$,
$V^{I}$. We know that it should coincide with our initial PB
of the F type.

 According to the Dirac procedure, for any functional
$F(n)$  we should find such linear combination 

$$\int m^{s}(X) g_{s}(X) dX $$
of the  constraints:

$$g_{k}(X) = V^{N+k}(X) - v_{k}(n(X)) \,\, , \,\,\,
k = 1,\dots,g $$
that the functional

$$F(n) + \int m^{s}(X) g_{s}(X) dX $$
leaves ${\cal M}^{N}$ invariant; i.e.

$$\{g_{k}(X), F(n) + \int m^{s}(Y) g_{s}(Y) dY \} = $$

$$= - \sum_{\nu=1}^{N} e_{\nu}
{\partial v_{k} \over \partial n^{\nu}} {d \over dX}
{\delta F \over \delta n^{\nu}(X) } + 
e_{N+k} {d \over dX} m^{k}(X) + $$

\begin{equation}
\label{uravsv}
+ \sum_{\nu=1}^{N} e_{\nu}
{\partial v_{k} \over \partial n^{\nu}} \left(
{\partial v_{s} \over \partial n^{\nu}}
{d \over dX} m^{s}(X) + \left( 
{\partial v_{s} \over \partial n^{\nu}}\right)_{X}
m^{s}(X) \right) \equiv 0
\end{equation}
The restricted bracket $\{\dots,\dots\}^{*}$ on
${\cal M}^{N}$ we define as

$$\{n^{\nu}(X), F(n)\}^{*} =
\{n^{\nu}(X),  F(n) + \int m^{s}(Y) g_{s}(Y) dY \} $$

 The relations (\ref{uravsv}) for $m^{s}(X)$ can be written
in the form

$$- \sum_{\nu=1}^{N} e_{\nu} 
{\partial v_{k} \over \partial n^{\nu}}
{d \over dX} {\delta F \over \delta n^{\nu}(X) } +
{d \over dX} \left(e_{N+k} \delta^{ks} +
\sum_{\nu=1}^{N} e_{\nu}
{\partial v_{k} \over \partial n^{\nu}}
{\partial v_{s} \over \partial n^{\nu}} \right)
m^{s}(X) - $$

$$- \sum_{\nu=1}^{N} e_{\nu}
\left( {\partial v_{k} \over \partial n^{\nu}} \right)_{X}
{\partial v_{s} \over \partial n^{\nu}} m^{s}(X) \equiv 0 $$

 We note now that the vectors ${\bf t}_{\nu}(n)$ tangent
to ${\cal M}^{N}$ can be written as 

\begin{equation}
\label{tbas}
{\bf t}_{\nu}(n) = (0,\dots,1,\dots,0,
{\partial v_{1} \over \partial n^{\nu}}, \dots ,
{\partial v_{g} \over \partial n^{\nu}})^{t}
\end{equation}
where $1$ stays at the position $\nu$. The vectors
${\bf q}_{k}(n)$ orthogonal to ${\cal M}^{N}$ can be chosen as

\begin{equation}
\label{qbas}
{\bf q}_{k}(n) = 
(e_{1} {\partial v_{k} \over \partial n^{1}}, \dots,
e_{N} {\partial v_{k} \over \partial n^{N}}, 0, \dots,
- e_{N+k}, \dots, 0)^{t}
\end{equation}
Therefore the quantities

\begin{equation}
\label{qqgks}
e_{N+k} \delta^{ks} +
\sum_{\nu=1}^{N} e_{\nu}
{\partial v_{k} \over \partial n^{\nu}}
{\partial v_{s} \over \partial n^{\nu}} =
({\bf q}_{k}(n), {\bf q}_{s}(n) ) = G_{ks}(n)
\end{equation}
are equal to the pairwise scalar products of the normal vectors
${\bf q}_{k}$.

 Let us introduce the inverse matrix $G^{ks}(n)$ such
that

$$G^{ks}(n) ({\bf q}_{s}(n), {\bf q}_{n}(n) )
= \delta^{k}_{n} $$
$k,n = 1,\dots,g$.

 It can be  easily shown that the quantities

\begin{equation}  
\label{omegasv}
\Omega^{k}_{n,\mu} = -
\sum_{\nu=1}^{N} e_{\nu} 
{\partial^{2} v_{n} \over \partial n^{\nu} \partial n^{\mu}}
{\partial v_{s} \over \partial n^{\nu}} G^{sk} = - G^{ks} 
({\bf q}_{s}, {\partial {\bf q_{n}} \over \partial n^{\mu}})
\end{equation}
are the connection coefficients of the normal connection for
${\cal M}^{N}$ following from the pseudo-Euclidean structure in
$R^{N+g}$  written in the basis $\{{\bf q}_{s}\}$ in the 
normal bundle.

 So we can write the relations (\ref{uravsv}) as

$$\nabla_{\perp} \left[ G_{kp} m^{p}(X) \right] -
\sum_{\nu=1}^{N} e_{\nu}
{\partial v_{k} \over \partial n^{\nu}}
\left( {\delta F \over \delta n^{\nu}(X) } \right)_{X}
\equiv 0 $$
where 

\begin{equation}
\label{nablacov}
\nabla_{\perp} = \delta^{s}_{k} {d \over dX} \pm
\Omega^{s}_{k,\mu} n^{\mu}_{X}
\end{equation}
is the covariant derivative with respect to $X$ applied to the 
lower(upper) indices in the normal bundle.

 From this we get

$$m^{k}(X) = G^{ks} \nabla_{\perp}^{-1} 
\sum_{\nu=1}^{N} e_{\nu}
{\partial v_{s} \over \partial n^{\nu}}
\left({\delta F \over \delta n^{\nu}(X) } \right)_{X} $$
We define the operator $\nabla_{\perp}^{-1}$ in a
"skew-symmetric" way as $\partial^{-1}$ above,
 i.e. $\nabla_{\perp}^{-1} \kappa_{s}$,
(where $\kappa_{s}(X) \rightarrow 0$ at 
$X \rightarrow \pm \infty$) is the sum of the 
solutions to the equation

$${d \over dX} \tau_{s} + \Omega^{t}_{s,\mu} n^{\mu}_{X}
\tau_{t} = \kappa_{s} $$
equal to $0$ at $-\infty$ and $+\infty$ respectively and 
divided by $2$.

 Besides that, we consider here the index $k$ in the formula

\begin{equation}
\label{prov}
\sum_{\nu=1}^{N} e_{\nu}
{\partial v_{k} \over \partial n^{\nu}}
\left({\delta F \over \delta n^{\nu}(X) } \right)_{X} 
\equiv q_{k}^{I}
\left({\delta F \over \delta n^{I}(X) } \right)_{X}
\end{equation}
 as a lower index 
in the basis $\{{\bf q}_{s}\}$

 So, for the bracket restricted to ${\cal M}^{N}$ we can write

$$\{n^{\nu}(X), F(n)\}^{*} = e_{\nu} {d \over dX}
{\delta F \over \delta n^{\nu}(X)} - 
e_{\nu} \left( {\partial v_{s} \over \partial n^{\nu}}
G^{sk} \nabla_{\perp}^{-1} 
\sum_{\mu=1}^{N} e_{\mu}
{\partial v_{k} \over \partial n^{\mu}}
\left( {\delta F \over \delta n^{\mu}(X)} \right)_{X}
\right)_{X} $$

 After some elementary calculations (using the 
expressions for operators $d/dX \nabla_{\perp}^{-1}$
and $\nabla_{\perp}^{-1} d/dX$ following from the form
(\ref{nablacov}) and the relations (\ref{qqgks}) and
(\ref{omegasv})) we can write this bracket in the form:

$$\{n^{\nu}(X), F(n)\}^{*} = e_{\nu} {d \over dX}
{\delta F \over \delta n^{\nu}(X)} - 
\sum_{\mu=1}^{N} e_{\nu}
{\partial v_{s} \over \partial n^{\nu}}
G^{sk} e_{\mu} {\partial v_{k} \over \partial n^{\mu}}
{d \over dX} {\delta F \over \delta n^{\mu}(X)} - $$

$$- e_{\nu} \left( \nabla_{\perp} G^{sk}
{\partial v_{s} \over \partial n^{\nu}} \right)
\sum_{\mu=1}^{N} e_{\mu}
{\partial v_{k} \over \partial n^{\mu}}
{\delta F \over \delta n^{\mu}(X)} + $$

\begin{equation}
\label{nvres}
+ \sum_{\mu=1}^{N}
\left( \nabla_{\perp} G^{sk} e_{\nu}
{\partial v_{s} \over \partial n^{\nu}} \right)
\nabla_{\perp}^{-1} 
\left( \nabla_{\perp} e_{\mu}
{\partial v_{k} \over \partial n^{\mu}} \right)
{\delta F \over \delta n^{\mu}(X)}
\end{equation}
Here  differential operators $\nabla_{\perp}$ act only
on the functions staying within the same brackets $(\dots)$.
It is necessary to differ their actions on upper and 
lower indices. So the expressions
$(\nabla_{\perp} G^{sk} e_{\nu} \partial v_{k}/\partial n^{\nu})$
and $(\nabla_{\perp} e_{\mu} \partial v_{s}/\partial n^{\mu})$
are just $(\nabla_{\perp} G^{sk} {\bf q}_{k})^{\nu}$ and
$(\nabla_{\perp} {\bf q}_{s})^{\mu}$ where we take 
$\nu, \mu = 1,\dots,N$ for their components.

 Formula (\ref{nvres}) gives us the restriction of the 
bracket (\ref{brev}) to any submanifold 
${\cal M}^{N} \subset R^{N+g}$ in the coordinates $n^{\nu}$,
$V^{I}$. Suppose now that ${\cal M}^{N}$ has a flat normal 
connection. This means that locally there exists such 
non-degenerate matrix $S^{k}_{n}(n)$ that the normal connection

\begin{equation}
\label{cancov}
\nabla_{\perp} = \left(S^{-1}\right)^{s}_{n} {d \over dX}
S^{k}_{s}
\end{equation}
- for the action on the covectors $\kappa_{k}$ and

\begin{equation}
\label{canvec}
\nabla_{\perp} = S^{n}_{s} {d \over dX}
\left(S^{-1}\right)^{s}_{k}
\end{equation}
for the action on the vectors $\xi^{k}$  in the normal bundle.

 We can take $S^{k}_{n}(U_{0}) = I$ and introduce the
basis of parallel vector fields

$${\bf N}_{n}(n) = S^{k}_{n}(n) {\bf q}_{k}(n) $$
where ${\bf q}_{k}(n)$ were defined in (\ref{qbas}).

 Since our normal connection preserves the scalar product
in $R^{N+g}$ we conclude that the pairwise scalar 
products of ${\bf N}_{k}(n)$ are constant at any $n$;
i.e.

$$({\bf N}_{k}, {\bf N}_{n}) = e_{N+k} \delta_{kn}$$
Therefore we have for the matrix $G^{kn}(n)$:

\begin{equation}
\label{canmet}
G^{kn}(n) = \sum_{s=1}^{g} e_{N+s} S^{k}_{s}(n) 
S^{n}_{s}(n)
\end{equation}

 Substituting (\ref{cancov}), (\ref{canvec}), (\ref{canmet})
and the expression corresponding to $\nabla_{\perp}^{-1}$ for
the covectors in (\ref{nvres}), we obtain after the simple
calculation:

$$\{n^{\nu}(X), F(n)\}^{*} = e_{\nu} {d \over dX}
{\delta F \over \delta n^{\nu}(X)} - $$

$$- \sum_{\mu=1}^{N} \sum_{q=1}^{g} e_{N+q} \left(
e_{\nu} {\partial v_{s} \over \partial n^{\nu}} S^{s}_{q}
\right) \left(e_{\mu} 
{\partial v_{k} \over \partial n^{\mu}} S^{k}_{q}\right)
{d \over dX} {\delta F \over \delta n^{\mu}(X)} - $$

$$- \sum_{\mu=1}^{N} \sum_{q=1}^{g} e_{N+q} \left(
{d \over dX} e_{\nu} {\partial v_{s} \over \partial n^{\nu}}
S^{s}_{q}\right) \left(e_{\mu} 
{\partial v_{k} \over \partial n^{\mu}} S^{k}_{q}\right)
{\delta F \over \delta n^{\mu}(X)} + $$

\begin{equation}
\label{flconbr}
+\sum_{\mu=1}^{N} \sum_{q=1}^{g} e_{N+q} \left(
{d \over dX} e_{\nu} {\partial v_{s} \over \partial n^{\nu}}
S^{s}_{q}\right) \left({d \over dX}\right)^{-1}
\left({d \over dX} e_{\mu}
{\partial v_{k} \over \partial n^{\mu}} S^{k}_{q}\right)
{\delta F \over \delta n^{\mu}(X)} 
\end{equation}
(The operator $(d/dX)^{-1}$ is defined as above in a 
"skew-symmetric" way.)

 So if we put:

$$f^{\nu}_{(k)}(n) \equiv 
e_{\nu} {\partial v_{s} \over \partial n^{\nu}}
S^{s}_{k}(n) = N^{\nu}_{k}(n) $$
for $\nu = 1,\dots,N$ (summation with respect to $s$) we
just obtain the expression corresponding to the bracket
(\ref{fercan}) where $f^{\nu}_{(k)}(U_{0}) = 0$. 
It is not hard  to see that the flows 
$N^{\nu}_{(k)X}$ correspond to the Weingarten operators, 
the expressions corresponding to $\delta^{\prime}$
and $\delta$ terms are equal to the restricted metric with upper
indices and to the corresponding connection respectively
in accordance with Ferapontov theorems.
So we proved the part (I) of the Theorem.

 Since $f^{\nu}_{(k)}(X) \rightarrow 0$ at 
$X \rightarrow \pm \infty$ on the space of rapidly decreasing 
 functions $n^{\nu}(X)$, we write 
$(d/dX)^{-1} f^{\nu}_{(k)X} \equiv f^{\nu}_{(k)}$. It is
easy to see that the functionals

$$\int n^{\nu}(X) dX $$
are  annihilators of the bracket (\ref{fercan}) on the
space of rapidly decreasing  functions.

 Consider now the functionals:

\begin{equation}
\label{hvham}
H_{n} = \int v_{n}(n) dX 
\end{equation}

 We have

$$\{n^{\nu}(X), H_{n}\}^{*} = e_{\nu} {d \over dX}
{\partial v_{n} \over \partial n^{\nu}} - $$

$$- \sum_{\mu=1}^{N} \sum_{q=1}^{g} e_{N+q} \left(
e_{\nu} {\partial v_{s} \over \partial n^{\nu}} S^{s}_{q}
\right) \left(e_{\mu}
{\partial v_{k} \over \partial n^{\mu}} S^{k}_{q}\right)
{d \over dX}
{\partial v_{n} \over \partial n^{\mu}} - $$

$$- \sum_{\mu=1}^{N} \sum_{q=1}^{g} e_{N+q} \left(
{d \over dX} e_{\nu} {\partial v_{s} \over \partial n^{\nu}}
S^{s}_{q}\right) \left(e_{\mu}
{\partial v_{k} \over \partial n^{\mu}} S^{k}_{q}\right)
{\partial v_{n} \over \partial n^{\mu}} + $$

$$+ \sum_{\mu=1}^{N} \sum_{q=1}^{g} e_{N+q} \left(
{d \over dX} e_{\nu} {\partial v_{s} \over \partial n^{\nu}}
S^{s}_{q}\right) \left({d \over dX}\right)^{-1}
\left({d \over dX} e_{\mu}
{\partial v_{k} \over \partial n^{\mu}} S^{k}_{q}\right)
{\partial v_{n} \over \partial n^{\mu}} $$

 We can write

$$\sum_{\mu=1}^{N} \left(e_{\mu}
{\partial v_{k} \over \partial n^{\mu}} S^{k}_{q}\right)
{\partial v_{n} \over \partial n^{\mu}} = 
({\bf N}_{q}, {\bf q}_{n}) + N^{N+n}_{q} =
({\bf N}_{q}, {\bf q}_{n}) - e_{N+n} S^{n}_{q} $$

$$\sum_{\mu=1}^{N} \left(e_{\mu}
{\partial v_{k} \over \partial n^{\mu}} S^{k}_{q}\right)
{d \over dX} {\partial v_{n} \over \partial n^{\mu}} =
({\bf N}_{q}, {d \over dX}{\bf q}_{n}) $$

$$\sum_{\mu=1}^{N} \left(e_{\mu}
{\partial v_{k} \over \partial n^{\mu}} S^{k}_{q}\right)_{X}
{\partial v_{n} \over \partial n^{\mu}} = 
({d \over dX}{\bf N}_{q}, {\bf q}_{n}) - 
e_{N+n} \left(S^{n}_{q}\right)_{X} = $$

$$= (\nabla_{\perp} {\bf N}_{q}, {\bf q}_{n}) -
e_{N+n} \left(S^{n}_{q}\right)_{X} = - e_{N+n} 
\left(S^{n}_{q}\right)_{X}$$
since $\nabla_{\perp} {\bf N}_{q}$ is tangent to $M^{N}$.

 So we have

$$\{n^{\nu}(X), H_{n}\}^{*} = {d \over dX} \left(
q^{\nu}_{n} - \sum_{q=1}^{g} e_{N+q} N_{q}^{\nu}
({\bf N}_{q}, {\bf q}_{n}) \right) + $$

$$+ \sum_{q=1}^{g} 
\left(
{d \over dX} e_{\nu} {\partial v_{s} \over \partial n^{\nu}}
S^{s}_{q}\right) \left(S^{n}_{q} - 
\left({d \over dX}\right)^{-1} \left(S^{n}_{q}\right)_{X}
\right) $$
for $\nu = 1,\dots,N$. As we know, any
${\bf q}_{n}$ is orthogonal to ${\cal M}^{N}$ and 
$\{{\bf N}_{q}\}$ give the pseudo-orthonormal basis in 
the normal bundle. Therefore 
the first term in this expression is zero. As for the last term,
we know that $S^{n}_{q}(X) \rightarrow \delta^{n}_{q}$ at
$X \rightarrow \pm \infty$.  So we have

$$\left({d \over dX}\right)^{-1} \left(S^{n}_{q}\right)_{X}
= S^{n}_{q}(X) - 
{S^{n}_{q}(+\infty) + S^{n}_{q}(-\infty) \over 2} = 
S^{n}_{q}(X) - \delta^{n}_{q} $$
and

$$\{n^{\nu}(X), H_{n}\}^{*} = {d \over dX}
\left(e_{\nu} {\partial v_{s} \over \partial n^{\nu}}
S^{s}_{n}\right) = {d \over dX} f^{\nu}_{(n)}$$
on the space of rapidly decreasing 
functions $n^{\nu}(X)$. So we proved the parts (II) and (III)
of the Theorem.

{\hfill Theorem is proved.}

\vspace{0.5cm}

 It is easy to see that the nonlocal tail of the MF
bracket (\ref{mokfer}) has the same form in any coordinate 
system ${\tilde U}^{\nu} = {\tilde U}^{\nu}(U)$. So if we 
require $f^{\nu}(0) \equiv 0$  in the case of 
the MF bracket, we obtain $f^{\nu}(n) \equiv n^{\nu}$
here.
 By  direct calculation we obtain the following:

\vspace{0.5cm}

{\bf Lemma 4.} {\it
For the variables

\begin{equation}
\label{vdef}
v^{\nu}(X) = {\partial}^{-1} n^{\nu}(X)
\end{equation}
with the nonlocal operator ${\partial}^{-1}$ defined in
(\ref{d1}) we have  
for the Poisson brackets (\ref{fercan})

$$\{v^{\nu}(X), v^{\mu}(Y)\} = $$

$$= - \epsilon^{\nu} \delta^{\nu\mu} \nu (X-Y) +
\sum_{k=1}^{g} e_{k} f^{\nu}_{(k)}(v_{X}) \nu (X-Y)
f^{\mu}_{(k)}(v_{Y}) = $$

\begin{equation}
\label{intform}
= - \sum_{\lambda = 1}^{N} \epsilon^{\lambda}
\delta^{\nu}_{\lambda} \nu (X-Y) \delta^{\mu}_{\lambda} +
\sum_{k=1}^{g} e_{k} f^{\nu}_{(k)}(v_{X}) \nu (X-Y)
f^{\mu}_{(k)}(v_{Y})
\end{equation}  
}

\vspace{0.5cm}

 So, the local and nonlocal parts of (\ref{fercan}) can
be unified after the non-local transformation (\ref{vdef}). 
The flows

$$v^{\nu}_{t_{\mu}} = \delta^{\nu}_{\mu} $$
and

$$v^{\nu}_{T_{k}} = f^{\nu}_{(k)}(v_{X}) $$
 commute with each other.   They  preserve  the bracket
(\ref{intform})  as follows from the general
theorem.

 For the MF-bracket we have in this case  

$$\{v^{\nu}(X), v^{\mu}(Y)\} = $$

$$= - \epsilon^{\nu} \delta^{\nu\mu} \nu (X-Y) +
c v^{\nu}_{X} \nu (X-Y) v^{\mu}_{Y} $$

\vspace{0.5cm}

 We would like to emphasize here 
that the  annihilators (Casimirs) of the bracket
(\ref{fer}) and the Hamiltonians for the flows
$w^{\nu}_{(k)\mu}U^{\mu}_{X}$ strongly depend on the boundary
conditions for the functions $U^{\nu}(X)$. 
We can see that it is not possible to divide a set of
$N+g$ "Canonical forms" (the restrictions of Euclidean coordinates
to ${\cal M}^{N}$) to the Casimirs of the bracket and
Hamiltonians for the flows from nonlocal tail until we fix
a point $y \in {\cal M}^{N}$ and define the corresponding loop 
space

$$L({\cal M},y) = \{\gamma : R^{1} \rightarrow {\cal M}^{N} :
\gamma(-\infty) = \gamma(+\infty) = y \in {\cal M}^{N}\}$$
corresponding to any point $y \in {\cal M}^{N}$. In this
case any global function $h(U)$ on ${\cal M}^{N}$ such that
$h(y)  = 0$ gives a "hydrodynamic type Hamiltonian"  

$$H = \int h(\gamma(X)) dX $$
on the loop space $L({\cal M},y)$ and the Hamiltonian flow
on $L({\cal M},y)$ defined in the invariant way on ${\cal M}^{N}$.
 
\vspace{0.5cm}

Remark.

 If we consider the Poisson bracket $\{\dots,\dots\}_{y}$
on the loop space $L({\cal M},y)$ for the Dirac restriction
(\ref{nvres}) of $e_{I}\delta^{IJ}d/dX$ on the general space   
${\cal M}^{N}$ in $R^{N+g}$ it is possible to introduce
the relations (\ref{cancov})-(\ref{canvec}) with some
matrix $S^{k}_{n}[\gamma](X)$ along every curve
$\gamma \in L({\cal M},y)$. The relations (\ref{flconbr})
can be also formally written on the space $L({\cal M},y)$,
but in this case the expressions  $S^{k}_{n}[\gamma](X)$
are nonlocal functionals. 
They are defined on the loop spaces $\gamma \in L({\cal M},y)$
 and depend
on the whole curve $\gamma$. However,
in this case it can be shown by the same way that the
integrals (\ref{ferann}) are still the local Casimirs for 
the bracket (\ref{nvres});  the functionals (\ref{hvham})  
generate the nonlocal flows

$$n^{\nu}_{T_{n}} = {1 \over 2} \sum_{q=1}^{g} {d \over dX}
\left(e_{\nu} {\partial v_{s} \over \partial n^{\nu}}
S^{s}_{q}[\gamma](X)\right) \left(S^{n}_{q}(-\infty) +
S^{n}_{q}(+\infty)[\gamma]\right) $$
on $L({\cal M},y)$.

\vspace{0.3cm}

 Now we prove that the general F-bracket with nondegenerate
$g^{\nu\mu}(U)$ has exactly $N$ Casimirs 
on the space $L({\cal M}^N,y)$
with fixed  point $y \in {\cal M}^{N}$.

\vspace{0.5cm}

{\bf Theorem 4.} {\it
Suppose we have a manifold ${\cal M}^{N}$ with non-degenerate
$g^{\nu\mu}(U)$ and the relations (\ref{metwein})-(\ref{rtenw})
for the affinors set $w^{\nu}_{(k)\mu}$. Then:

  For any point $y \in {\cal M}^{N}$ there exist locally
exactly $N+g$ linearly independent functions $V^{I}_{y}(U)$,
$I = 1,\dots,N+g$ such that the functionals

$$H^{I} = \int V^{I}_{y}(U) dX $$
generate the flows proportional to
$w^{\nu}_{(k)\mu}(U) U^{\mu}_{X}$
on the space $L({\cal M},y)$ of loops close enough to
$y$. The functions $V^{I}_{y}(U)$ can be chosen in such a way
that for close enough $y$, $y^{\prime}$

\begin{equation}
\label{mmprime}
V^{I}_{y}(U) = A^{I}_{J}(y,y^{\prime})
V^{J}_{y^{\prime}}(U) + B^{I}(y,y^{\prime})
\end{equation}
with some constants $B^{I}(y,y^{\prime})$ and
the matrix $A^{I}_{J}(y,y^{\prime})$ orthogonal with respect
to diagonal metric

$$g_{IJ} =
\epsilon_{I} \delta_{IJ},\,\,\,\,\, I = 1,\dots,N+g  $$
where $\epsilon^{I} = \epsilon^{I},\,\, I = 1,\dots,N$,
$\epsilon^{I} = e_{I-N},\,\, I = N+1,\dots,N+g$.
}

\vspace{0.5cm}

Proof.

 We should prove that the linear space of functions
$F_{y}(n)$ such that: $F_{y}(0) \equiv 0$,

\begin{equation}
\label{fgkx}
{\partial F_{y} \over \partial n^{\mu}}
f^{\mu}_{(k)X} \equiv G_{y(k)X}
\end{equation}
for some $G_{y(k)}(U)$, $G_{y(k)}(y) \equiv 0$ and

\begin{equation}
\label{fgder}
{d \over dX} \left(\epsilon^{\nu}
{\partial F_{y} \over \partial n^{\nu}} -
\sum_{k=1}^{g} e_{k} \left(f^{\nu}_{(k)} f^{\mu}_{(k)}
{\partial F_{y} \over \partial n^{\mu}} -
f^{\nu}_{(k)} G_{y(k)}\right) \right) =
\sum_{k=1}^{g} \alpha_{k} f^{\nu}_{(k)X}
\end{equation}
(any $\alpha_{1},\dots,\alpha_{g}$) is at most
$N+g$ - dimensional ($\epsilon^{\nu} = \pm 1$). We have

$$\epsilon^{\nu}
{\partial F_{y} \over \partial n^{\nu}} -
\sum_{k=1}^{g} \left(f^{\nu}_{(k)} f^{\mu}_{(k)}
{\partial F_{y} \over \partial n^{\mu}} - 
f^{\nu}_{(k)} G_{y(k)}\right) =
\sum_{k=1}^{g} \alpha_{k} f^{\nu}_{(k)} + \beta^{\nu} $$
for some constants $\beta^{1},\dots,\beta^{N}$, and

$$\sum_{k=1}^{g} f^{\mu}_{(k)} 
{\partial F_{y} \over \partial n^{\mu}} \left(
\delta^{sk} - e_{k} \sum_{\nu=1}^{N} \epsilon^{\nu}
f^{\nu}_{(s)} f^{\nu}_{(k)} \right) +
\sum_{k=1}^{g} e_{k} \left( \sum_{\nu=1}^{N}
\epsilon^{\nu} f^{\nu}_{(s)} f^{\nu}_{(k)} G_{y(k)}
\right) = $$

$$= \sum_{k=1}^{g} \alpha_{k} \sum_{\nu=1}^{N}
\epsilon^{\nu} f^{\nu}_{(s)} f^{\nu}_{(k)} +
\sum_{\nu=1}^{N} \epsilon^{\nu} f^{\nu}_{(s)} \beta^{\nu} $$

 For the points close enough to $y$ the matrix

$$\delta^{sk} - e_{k} \sum_{\nu=1}^{N} \epsilon^{\nu}
f^{\nu}_{(s)} f^{\nu}_{(k)}$$
is non-degenerate, and we can locally express the value
$f^{\mu}_{(k)}\partial F_{y}/\partial n^{\mu}$  as a
function of $G_{y(s)}$, $f^{\nu}_{(s)}$, $\alpha_{k}$,
$\beta^{\nu}$ at every point. After that we obtain the
equations:

\begin{equation}
\label{felg}
{\partial F_{y} \over \partial n^{\nu}} =
\epsilon^{\nu} \Lambda^{\nu}(G_{y(s)},f^{\mu}_{(s)},
\alpha_{k},\beta^{\nu})
\end{equation}
and the linear (non-homogeneous) equations  

$${\partial G_{y(k)} \over \partial n^{\nu}} =
\sum_{\mu=1}^{N} \epsilon^{\mu}
{\partial f^{\mu}_{(k)} \over \partial n^{\nu}}
\Lambda^{\mu}(G_{y(s)},f^{\mu}_{(s)},
\alpha_{k},\beta^{\nu}) $$
for every derivative of $G_{y(k)}$ with the normalizing
conditions $G_{y(k)}(y) = 0$ which give us the unique
$G_{y(k)}(U)$ at every $\{\alpha_{k},\beta^{\nu}\}$. So,
from (\ref{felg}) we obtain that the family $F_{y}(U)$ is
at most $N+g$ parametric.

 The functions constructed in the proof of Theorem 3 then
give us $N$ linearly independent densities of annihilators
of the bracket $n^{\nu}(U)$ and $g$ functions $v_{k}(n)$
(generating linearly independent flows
$e_{k} f^{\nu}_{(k)X}$) which satisfy all the
conditions of the Theorem.

\hfill{Theorem is proved.}

\vspace{0.5cm} 

 The construction of  the Dirac restriction for the F-brackets
leads also to the following statement:

\vspace{0.5cm}

{\bf Theorem 5.}{\it

The symplectic form for any F-bracket with
nondegenerate tensor $g^{\nu\mu}(U)$ is weakly
nonlocal and can be written in the form

\begin{equation}
\label{sympfer}
\Omega_{\nu\mu}(X,Y) = \sum_{I=1}^{N+g} \epsilon_{I}
{\partial V_{I} \over \partial U^{\nu}}(X)
\nu (X-Y) {\partial V_{I} \over \partial U^{\mu}}(Y)
\end{equation}
where $V_{I}(U)$ are the restrictions of the Euclidean
coordinates to ${\cal M}^{N}$ (canonical forms)
and $\epsilon_{I} = \pm 1$ according to the signature
of the flat metric in $R^{N+g}$.}\footnote{The symplectic form 
for the case of the MF-bracket was considered also by M.V.Pavlov 
(private communication).}

\vspace{0.5cm}

Proof. For the proof of the Theorem we just note
that the symplectic form of the F-bracket should coincide with 
the restriction of the symplectic form for the corresponding
DN-bracket in $R^{N+g}$. The symplectic form for the DN-bracket
is given by

$$\Omega_{IJ}(X,Y) = \epsilon_{I} \delta_{IJ} \nu(X-Y) $$
and it it easy to see that its restriction to ${\cal M}^{N}$
is given by the formula (\ref{sympfer}).

{\hfill Theorem is proved.}

\vspace{0.5cm}

 At the end we give some classification for the special
class of  brackets (\ref{fercan}) which represent the
multi-parametric Poisson pencils.

\vspace{0.5cm}

{\bf Theorem 6.} {\it
I. The expression

$$\{n^{\nu}(X), n^{\mu}(Y)\} = \left( \epsilon^{\nu}
\delta^{\nu\mu} - \sum_{k=0}^{g} {\alpha}_{k} f^{\nu}_{(k)}(n)
f^{\mu}_{(k)}(n) \right)
\delta^{\prime}(X-Y) - $$
\begin{equation}
\label{pencil}
- \sum_{k=0}^{g} {\alpha}_{k}  
\left( f ^{\nu}_{(k)}(n) \right)_{X} f^{\mu}_{(k)}(n)
\delta (X-Y) + \sum_{k=0}^{g} {\alpha}_{k}
\left( f ^{\nu}_{(k)}(n) \right)_{X} \nu (X-Y)
\left( f ^{\mu}_{(k)}(n) \right)_{Y}
\end{equation}
with a linearly independent set of $f ^{\nu}_{(k)}(n)$
defines a Poisson bracket at any 
$({\alpha}_{1},\dots,{\alpha}_{N})$ if and only if:

1) The flows 

$$n^{\nu}_{T_{k}} = \left( f ^{\nu}_{(k)} \right)_{X}$$
are Hamiltonian with respect to local Poisson bracket

$$\{n^{\nu}(X), n^{\mu}(Y)\}_{0} =
\epsilon^{\nu} \delta^{\nu\mu} \delta^{\prime}(X-Y) $$
with some local Hamiltonian functions $H_{k}$, i.e. there 
exist such functions $h_{k}(n)$ that

$$f ^{\nu}_{(k)}(n) \equiv \epsilon^{\nu}
{\partial h_{k} \over \partial n^{\nu}}$$

2) The Hamiltonians

$$H_{k} = \int h_{k}(X) dX $$
commute with each other with respect to the bracket 
$\{\dots,\dots\}_{0}$ and moreover:

$$\{h_{k}, H_{k^{\prime}}\}_{0} = 
\tau_{kk^{\prime}}(h_{k})h_{kX} =
\left(W_{kk^{\prime}}(h_{k})\right)_{X}$$
for some $\tau(h_{k})$ and $W(h_{k})$.

\vspace{0.5cm}

II. If the conditions of (I) take place (i.e. we have a
Poisson pencil) then the functionals $H_{k}$ generate local
Hamiltonian flows and commute with each other with respect to
the full bracket (\ref{pencil}). Besides that, the maximal 
functionally independent subset of $h_{k}(n)$ defines 
at any $(\alpha_{1},\dots,\alpha_{g})$ a 
closed sub-bracket on the space of corresponding $h_{k}(X)$.
}

\vspace{0.5cm}

  Proof. 

  First of all we note that our bracket at any 
$(\alpha^{1},\dots,\alpha^{g})$ can be written in the
Canonical form (\ref{fercan}) after the linear
changes of the functions $f^{\nu}_{(k)}(n)$. So we
can use here the results of Theorem 2  replacing
the quantities $e_{s}$ by $\alpha_{s}$ in every case.
After that from the constant 
part of the  condition $(2)$ of Theorem 2 we get the equation

$$\epsilon^{\mu} 
{\partial f^{\nu}_{(k)} \over \partial n^{\mu}} =
\epsilon^{\nu}
{\partial f^{\mu}_{(k)} \over \partial n^{\nu}} $$
So locally we have:

\begin{equation}
\label{fh}
f^{\nu}_{(k)} \equiv \epsilon^{\nu} 
{\partial h_{k} \over \partial n^{\nu}}
\end{equation}
for some $h_{k}(n)$. Therefore we are coming to the first 
statement of our Theorem. The commutativity of the functionals
$H_{k} = \int h_{k}(X) dX$ with respect to the bracket
$\{\dots,\dots\}_{0}$ now follows  from  condition
$(1)$ of  Theorem 2.

 Now from the linear term (with respect to $\alpha_{s}$)  of 
the condition $(2)$ of Theorem 2 we have:

$$\epsilon^{\nu} \epsilon^{\mu}
\sum_{\lambda} \epsilon^{\lambda}
{\partial^{2} h_{k} \over \partial n^{\lambda} \partial n^{\nu}}
{\partial h_{s} \over \partial n^{\lambda}}
{\partial h_{s} \over \partial n^{\mu}} =
\epsilon^{\mu} \epsilon^{\nu}
\sum_{\lambda} \epsilon^{\lambda}
{\partial^{2} h_{k} \over \partial n^{\lambda} \partial n^{\mu}}
{\partial h_{s} \over \partial n^{\lambda}}
{\partial h_{s} \over \partial n^{\nu}} $$
for any $k$, $s$, $\nu$, $\mu$. This fact means that the rows
(with respect to $\nu$) 
${\partial h_{s} \over \partial n^{\nu}}$
and 
$\sum_{\lambda} \epsilon^{\lambda}
{\partial^{2} h_{k} \over \partial n^{\lambda} \partial n^{\nu}}
{\partial h_{s} \over \partial n^{\lambda}}$ are linearly 
dependent for any $k$ and $s$, i.e.

\begin{equation}
\label{lindep}
\sum_{\lambda} \epsilon^{\lambda}
{\partial^{2} h_{k} \over \partial n^{\lambda} \partial n^{\nu}}
{\partial h_{s} \over \partial n^{\lambda}} =
\tau_{sk}(n) 
{\partial h_{s} \over \partial n^{\nu}}
\end{equation}

 After  multiplying (\ref{lindep}) by $n^{\nu}_{X}$ and the
summation with respect to $\nu$, we obtain the following equality

$${d h_{s} \over d T_{k}} = \tau_{sk}(n) h_{sX} $$
This expression should be equal to some $(W_{sk}(n))_{X}$ because
of the commutativity of $H_{s}$ and $H_{k}$ with respect to
the bracket $\{\dots,\dots\}_{0}$. So we conclude that 
$W_{sk}(n)$ can be expressed  in terms of the values of
$h_{s}(n)$:

\begin{equation}
\label{tauhs}
\tau_{sk}(n) \equiv \tau_{sk}(h_{s}) 
\end{equation}
and $W_{sk}(\xi) = \int \tau_{sk}(\xi) d\xi$.

  Therefore the requirements of our Theorem are equivalent
to the requirements corresponding to the analogs of $(1)$ and
$(2)$ of Theorem 2 for all $(\alpha_{1},\dots,\alpha_{g})$.
It is not hard to check by direct calculation that 
the analog of the condition $(3)$ of Theorem 2 for any
$(\alpha_{1},\dots,\alpha_{g})$ follows also from (\ref{fh}),
(\ref{lindep}) and $(\ref{tauhs})$. So we proved  part $(I)$
of the Theorem.

 To prove  part $(II)$ we note that all $H_{s}$ generate
the local flows according to the full bracket (\ref{pencil}) since
they are the integrals of the flows 
$w^{\nu}_{(k)\mu} n^{\mu}_{X}$ from the non-local part of a 
bracket according to the commutativity of the $\{H_{k}\}$
with respect to the bracket $\{\dots,\dots\}_{0}$.
To prove the rest of the Theorem we mention that we deduce
from (\ref{lindep}) and (\ref{tauhs}) the relation

$$\sum_{\nu} {\partial h_{k} \over \partial n^{\nu}}
\epsilon^{\nu} 
\left( {\partial h_{s} \over \partial n^{\nu}} \right)_{X} =
\left( W_{sk}(h_{k}) \right)_{X} $$
We can obviously choose $W_{sk}(h_{k})$ such that

$$\sum_{\nu} {\partial h_{k} \over \partial n^{\nu}}
\epsilon^{\nu} {\partial h_{s} \over \partial n^{\nu}} =
W_{sk}(h_{k}) + W_{ks}(h_{s}) $$

 So for the bracket $\{h_{s}(X),h_{t}(Y)\}$ 
after the simple calculation we can write the 
the following equalities

$$\{h_{s}(X),h_{t}(Y)\} = \left( W_{st}(h_{s}) + W_{ts}(h_{t})
\right) \delta^{\prime}(X-Y) -$$

$$ - \left( \sum_{k=1}^{g}
\alpha_{k} \left( W_{sk}(h_{s}) + W_{ks}(h_{k}) \right)
\left( W_{kt}(h_{k}) + W_{tk}(h_{t}) \right) \right)
\delta^{\prime}(X-Y) + $$

$$+ \left(W_{st}(h_{s}) \right)_{X}
\delta(X-Y) - \sum_{k=1}^{g} \alpha_{k} 
\left( W_{sk}(h_{s}) W_{kt}(h_{k}) \right)_{X} \delta(X-Y) -$$

$$- \sum_{k=1}^{g} \alpha_{k} \left(W_{ks}(h_{k}) 
\left( W_{kt}(h_{k})\right)_{X} +
\left( W_{sk}(h_{s}) \right)_{X} W_{tk}(h_{t}) \right)
\delta(X-Y) +$$

\begin{equation}
\label{subbr}
+ \sum_{k=1}^{g} \alpha_{k} 
\left( W_{sk}(h_{s}) \right)_{X} \nu(X-Y) 
\left( W_{tk}(h_{t}) \right)_{Y}
\end{equation}
where all $W_{kk^{\prime}}$ are expressed in terms of the
corresponding $h_{k}$ and

$$\{h_{s}(X),H_{t}\} = \left( W_{st}(h_{s}) \right)_{X} -
\sum_{k=1}^{g} \alpha_{k} 
\left( W_{sk}(h_{s}) W_{kt}(h_{k}) - 
Z_{kst}(h_{k}) \right)_{X} $$
where $Z_{kst}(\xi) = \int W_{ks}(\xi) 
W^{\prime}_{kt}(\xi) d\xi$.

 So we conclude that any $H_{s}$ and $H_{t}$ commute with 
each other with respect to the full pencil (\ref{pencil}), and 
the maximal functionally independent subset of $\{h_{k}\}$
form a sub-bracket (\ref{subbr}) at any 
$(\alpha_{1},\dots,\alpha_{g})$.

{\hfill Theorem is proved.}

\vspace{0.5cm}

{\bf Simple example.} Let us consider the bracket

$$\{n^{\nu}(X), n^{\mu}(Y)\}_{0} =
\epsilon^{\nu} \delta^{\nu\mu} \delta^{\prime}(X-Y) $$

 It's easy to show that any set of functions $h_{k}(R^{2})$
where $R^{2} = \sum_{\nu} \epsilon^{\nu} n^{\nu} n^{\nu}$
satisfy the requirements of Theorem 3 as the Hamiltonian 
functions for the flows $(f^{\nu}_{(k)})_{X}$. So, any
linearly independent set of such functions gives us the
Poisson pencil (\ref{pencil}) according to the relations

$$f^{\nu}_{(k)}(n) = \epsilon^{\nu}
{\partial h_{k} \over \partial n^{\nu}} =
2 h^{\prime}_{k}(R^{2}) n^{\nu} $$

 In particular, if we take only one function
$h(R^{2}) = R^{2}/2$ we obtain the Canonical form of the
MF bracket corresponding to the constant
curvature $\alpha$.

\end{document}